\documentclass[aps,prd,showpacs,notitlepage,nofootinbib,superscriptaddress,floatfix,showkeys,twocolumn]{revtex4-1}
\usepackage{comment}
\usepackage{graphicx}
\usepackage{amsmath}
\usepackage{float}
\usepackage{subfigure}
\usepackage{tensor}
\usepackage{natbib}
\usepackage[usenames]{color}
\usepackage{amssymb}
\usepackage{multirow}
\usepackage{longtable}
\usepackage{lipsum, babel}
\usepackage{epstopdf}
\usepackage{multirow}
\newcommand{\bea}{\begin{aligned}}
\newcommand{\eea}{\end{aligned}}
\newcommand{\be}{\begin{equation}}
\newcommand{\ee}{\end{equation}}

\newcommand{\nno}{\nonumber}
\newcommand{\bse}{\begin{subequations}}
\newcommand{\ese}{\end{subequations}}

\usepackage{hyperref}
\hypersetup{
     colorlinks   = true,
     citecolor    = red,
     linkcolor    = blue,
     urlcolor     = blue,
}

\newcommand{\bmm}{\begin{multline}}
\newcommand{\emm}{\end{multline}}

\def\beq{\begin{equation}}
\def\eeq{\end{equation}}
\def\bea{\begin{eqnarray}}
\def\eea{\end{eqnarray}}
\def\be{\begin{equation}}
\def\ee{\end{equation}}

\def\nno{}
\def\bse{\begin{subequations}}
\def\ese{\end{subequations}}

\def\aend{a_{\rm end}}

\def\are{a_{\rm re}}
\def\Are{A_{\rm re}}
\def\aend{a_{\rm end}}
\def\tre{T_{\rm re}}

\def\wphi{w_{\rm \phi}}

\def\ns{n_{\rm s}}
\def\tmax{T_{\rm max}}
\def\tbbn{T_{\rm BBN}}
\def\mphi{m_{\rm \phi}}

\numberwithin{equation}{section}
\begin{document}
\title{Constraining Reheating Temperature, Inflaton-SM Coupling and Dark Matter Mass in Light of ACT DR6 Observations}
\author{Rajesh Mondal}
\email{mrajesh@iitg.ac.in}
\affiliation{Department of Physics, 
Indian Institute of Technology Guwahati, Guwahati 781039, Assam, India}%
\author{Sourav Mondal}%
\email{sm206121110@iitg.ac.in}
\affiliation{Department of Physics, 
Indian Institute of Technology Guwahati, Guwahati 781039, Assam, India}
\author{Ayan Chakraborty}%
\email{chakrabo@iitg.ac.in}
\affiliation{Department of Physics, 
Indian Institute of Technology Guwahati, Guwahati 781039, Assam, India}
\pagenumbering{arabic}
\renewcommand{\thesection}{\arabic{section}}

\begin{abstract}
We explore the phenomenological implications of the latest Atacama Cosmology Telescope (ACT) DR6 observations, in combination with Planck 2018, BICEP/Keck 2018, and DESI, on the physics of inflation and post-inflationary reheating. We focus on the $\alpha$-attractor class of inflationary models (both E- and T-models) and consider two reheating scenarios: perturbative inflaton ($\phi$) decay ($\phi \rightarrow b\,b$) and inflaton annihilation ($\phi\, \phi \rightarrow b\,b$) into Standard Model (SM) bosonic particles ($b$). By solving the Boltzmann equations, we derive bounds on key reheating parameters, including the reheating temperature, the inflaton equation of state (EoS), and the inflaton-SM coupling in light of ACT data. To accurately constrain the coupling, we incorporate the Bose enhancement effect in the decay width. To ensure the validity of our perturbative approach, we also identify the regime where nonperturbative effects, such as parametric resonance, become significant. Additionally, we include indirect constraints from primordial gravitational waves (PGWs), which can impact the effective number of relativistic species, $\Delta N_{\rm eff}$. These constraints further bound the reheating temperature, particularly in scenarios with a stiff EoS. Finally, we analyze dark matter (DM) production through purely gravitational interactions during reheating and determine the allowed mass ranges consistent with the constrained reheating parameter space and recent ACT data. 
\end{abstract}

\maketitle
\newpage
\section{\textbf{Introduction}}
Over the past few decades, high-precision measurements of the cosmic microwave background (CMB) have marked the beginning of a new era in precision cosmology. Cosmic Inflation, a widely accepted paradigm of the early universe, was originally proposed to address several fundamental issues of the standard Big Bang model, such as the horizon and flatness problems \cite{Guth:1982ec,Starobinsky:1982ee,Albrecht:1982mp}. Beyond these successes, inflation also provides a natural mechanism for generating the primordial fluctuations that seed the formation of large-scale structure. These primordial fluctuations evolve over time and eventually manifest as the anisotropies observed in the CMB \cite{Sofue:2000jx,CMB-S4:2016ple,Planck:2018jri}, which are quantitatively described by key observables such as the scalar spectral index $\ns$ and the tensor-to-scalar ratio $r$. The Planck 2018 \cite{Planck:2018jri} and BICEP/Keck Array 2018 \cite{BICEP:2021xfz,BICEP2:2018kqh} (BK18) observations gives constraints on $\ns = 0.9649 \pm 0.004$ and $r\leq0.036$ at $95\%$ confidence level (C.L). More recently, updated measurements from the Atacama Cosmology Telescope (ACT) \cite{ACT:2025fju,ACT:2025tim} leave $r$ largely unchanged, but suggest a slightly higher value of $\ns$ compared to the original Planck results. A combined analysis of Planck and ACT data (referred to as P+ACT) yields $\ns = 0.9709 \pm 0.0038$. When further supplemented with CMB lensing and Baryon Acoustic Oscillation (BAO) data from the Dark Energy Spectroscopic Instrument (DESI), as a combination P+ACT+LB+BK18-the scalar spectral index is found to be $\ns = 0.9743 \pm 0.0034$ \cite{ACT:2025fju,ACT:2025tim}. This value deviates from the original Planck estimate by approximately $2\sigma$, potentially indicating a shift in our understanding of early-universe physics. In this context, the latest observations have prompted a re-evaluation of several inflationary models to check their consistency with the new ACT data \cite{Kallosh:2025,Aoki:2025, Berera:2025,Dioguardi:2025a,Gialamas:2025,Salvio:2025,Antoniadis:2025,Kim:2025,Dioguardi:2025b,Gao:2025,He:2025,Pallis:2025,Drees:2025,Haque:2025uri, Zharov:2025evb,Liu:2025qca,Haque:2025uis,Yogesh:2025wak,Khan:2022odn,Gangopadhyay:2022vgh,Gialamas:2025ofz,Maity:2025czp,Wolf:2025ecy}.

However, inflation is not the end of the story. The universe must transition from the inflationary phase to a hot, radiation-dominated era- a process known as reheating. During this phase, the energy stored in the inflaton field is transferred to Standard Model (SM) particles, ultimately leading to the thermal universe after inflaton domination. Understanding this transition is crucial for unveiling the microphysical properties of the inflaton and its interactions with the visible sector. Reheating is typically characterized by two key parameters: the reheating temperature $\tre$, and the effective EoS parameter $\wphi$ that governs the dynamics of the inflaton field during this epoch. Both parameters are related to the inflaton potential $V(\phi)$ and its coupling to SM particles. In this study, we consider reheating scenarios in which the inflaton condensate either decays or annihilates into SM bosonic particles, through the channels $\phi \rightarrow b\,b$ (bosonic decay) and $\phi\,\phi\rightarrow {b\, b}$ ( bosonic annihilation). The produced particles, assumed to be massless, are expected to thermalize rapidly, thereby creating a thermal bath. Due to the absence of direct observational probes of reheating, the reheating temperature spans a wide range. There is a lower limit on the reheating temperature originating from BBN as $T_{\rm BBN} \simeq 4$ MeV \cite{Kawasaki:2000en,Hannestad:2004px,Barbieri:2025moq,deSalas:2015glj}, and an upper limit coming from the inflation scale as $T_{\rm re}\simeq 10^{15}$ GeV \cite{Planck:2018jri,BICEP:2021xfz}, which is called the instantaneous reheating temperature. Similarly, parameters like the reheating temperature and inflaton-SM coupling remain largely unconstrained due to the lack of direct observational evidence.

In recent years, significant progress has been made in constraining both inflationary and reheating parameters by combining BICEP/$Keck$ 2018 (BK18) data with Planck 2018 results \cite{Ellis:2021kad, Chakraborty:2023ocr}. In addition, several recent studies \cite{Drewes:2022nhu,Drewes:2023bbs,Liu:2025sut} have explored the potential of future CMB experiments, such as CMB-S4 \cite{CMB-S4:2020lpa,CMB-S4:2016ple}, LiteBIRD \cite{LiteBIRD:2022cnt}, and AliCPT \cite{Li:2017drr,Li:2018rwc} to constrain reheating parameters through improved sensitivity to inflationary observables. In this work, we extend such analyses by exploring constraints on the parameters of the $\alpha$-attractor inflationary models, along with key reheating parameters—including the reheating temperature $\tre$, the inflaton’s equation of state $w_{\phi}$, and the inflaton–SM coupling—using the latest ACT DR6 data in combination with Planck18, BK18, and DESI observations. In addition, we investigate the dark sector by constraining the masses of dark matter particles that could be produced during the post-inflationary reheating era. Consequently, the bounds obtained on reheating parameters also allow us to place indirect constraints on the viable DM mass ranges.

Alongside observational bounds, we also include two theoretical constraints. First, the overproduction of primordial gravitational waves in scenarios with stiff post-inflationary dynamics ($\wphi>1/3$) can lead to excessive contributions to the effective number of relativistic species $\Delta N_{\rm eff}$, which is tightly constrained by BBN and CMB data. This provides a lower bound on reheating temperature $\tre$. Second, our analysis assumes perturbative inflaton decay or annihilation. For large couplings, non-perturbative processes like preheating can dominate, making the perturbative approach invalid. We therefore impose an upper limit on the coupling strength to ensure the consistency of our framework. 

The paper is organized as follows: In Sec.\ref{sc2}, we begin by introducing the inflationary model. In Sec.\ref{sc3}, we provide a detailed discussion of the dynamics of reheating after inflation for two different production channels. In Sec.\ref{sc4}, we discuss the constraints on the reheating temperature and inflaton EoS arising from the overproduction of PGWs. In Sec.\ref{sc6}, we discuss the validity of the perturbative reheating framework and constrain the coupling parameter. Then, in Sec.\ref{sc7}, we discuss the final constraints on the inflationary and post-inflationary reheating parameters. After that, Sec.\ref{sc8} is dedicated to constraining the dark sector. Finally, we conclude in Sec.~\ref{sc9}.

\section{\textbf{ Inflationary Model}}\label{sc2}
As discussed in the introduction, we consider a class of single-field inflationary models known as $\alpha$-attractors, which include both E-models and T-models. The general form of the potential is given by
\be
\label{attractorpotential}
V(\phi)  \;=\; 
\begin{cases}
\Lambda^4\,\left[1-e^{-\sqrt{\frac{2}{3\,\alpha}}\phi/M_p}\right]^{2\,n}\,, & {\rm E - model}\,,\\[10pt]
\Lambda^4\,\tanh^{2\,n}\left(\frac{\phi}{\sqrt{6\alpha}\,M_p}\right)\,,&  {\rm T-model}\,, \\[10pt]
\end{cases}
\ee
where $\Lambda$ sets the overall scale of the potential, $M_p = 2.4 \times 10^{18}$ GeV is the reduced Planck mass, and $\alpha$, $n$ are dimensionless parameters that control the shape of the potential. These model parameters are constrained by inflationary observables encoded in the curvature perturbation power spectrum, 
\beq
\Delta^2_{\mathcal{R}}  =A_{\mathcal{R}} (k/k_0)^{n_s -1}
\eeq
where the amplitude $A_{\mathcal{R}}$ is measured to be $(2.19 \pm 0.06) \times 10^{-9}$ at the pivot scale $k_0 = 0.05~\text{Mpc}^{-1}$. In addition to scalar perturbations, tensor modes are characterized by a nearly scale-invariant tensor power spectrum $\Delta^2_{\mathcal{T}} = A_{\mathcal{T}}$, with its amplitude constrained via the tensor-to-scalar ratio $r = A_{\mathcal{T}} / A_{\mathcal{R}} \lesssim 0.038$ \cite{ACT:2025fju,ACT:2025tim}. The two key inflationary observables $(\ns, r)$ are expressed in terms of the potential slow-roll parameters $\epsilon$ and $\eta$ as
\be \label{inflationp}
 n_{s}= 1- 6 \epsilon(\phi)+ 2 \eta(\phi)~,~r=16\epsilon(\phi)\,,
\ee
where $\epsilon$ and $\eta$ are defined through derivatives of the potential $V(\phi)$,
\beq
\epsilon(\phi) = \frac{M_p^2}{2} \left( \frac{V'(\phi)}{V(\phi)} \right)^2\,, \quad
\eta(\phi) = M_p^2 \left( \frac{V''(\phi)}{V(\phi)} \right)\,.
\eeq

In any inflationary model, the key CMB observables $(n_s, r)$ are typically related to two fundamental inflationary quantities, the inflationary energy scale $H_k$, and the duration of inflation, i.e., inflationary e-folding number $N_k$. Under the slow roll approximation, all are defined as, $k=k_0=0.05\,\mbox{Mpc}^{-1}$ as, 
\begin{eqnarray}\label{Hk}
H_{\rm k} &=& \frac{\pi M_{\rm p}\sqrt{r\,A_{\mathcal R}}} {\sqrt{2}} \simeq  \sqrt{\frac{V(\phi_{\rm k})}{3 M_{\rm p}^2} } \\ 
N_{\rm k}&  =& \int_{\phi_{\rm k}}^{\phi_{\rm end}} \frac{|d\phi|}{\sqrt{2\epsilon(\phi)} M_{\rm p}} ~~.
\end{eqnarray} 
Here $\phi_k$ and $\phi_{\rm end}$ denote the inflaton field values at horizon crossing and at the end of inflation, respectively.
More general expressions for the $N_{\rm k}$ and $H_{\rm k}$ in terms of inflationary parameters can be obtained in terms of inflationary parameters (see \cite{Drewes:2017fmn, Garcia:2020wiy} for details). 
The tensor-to-scalar ratio $r$ turns out as (see, last expression of Eq.\ref{inflationp})
\be
\label{rattrac}
r\;=\;
\begin{cases}
\frac{16n^2}{3\alpha}\left(e^{\sqrt{\frac{2}{3\,\alpha}}\frac{\phi_k}{M_{\rm p}}}-1\right)^{-2}\,,& {\rm E- model},\\[10pt]
\frac{16n^2}{3\alpha} \text{csch}^2\left(\sqrt{\frac{2}{3\,\alpha}}\frac{ \phi _k}{M_{\rm p}}\right)\,,& {\rm T- model}.
\end{cases}
\ee 
Post-inflationary dynamics, which we call reheating, will be controlled by the energy of the inflaton ($\rho_{\phi}^{\rm end}$),
\be\label{ic}
\rho^{\rm end}_{\phi} 
\sim \frac{\Lambda^4}{\alpha_1^n} \left[\frac{\phi_{\rm end}}{M_{\rm p}}\right]^n = \frac{4.1 \times 10^{64}}{\alpha_1^n} \left[\frac{\phi_{\rm end}}{M_p}\right]^n \left[\frac{\Lambda}{5.8\times10^{-3}M_{\rm p}}\right]^4 ,
\ee
which is in unit of ${\rm GeV}^4$ and  $\alpha_1=(\sqrt{{3\alpha}/{2}}, \sqrt{6\alpha})$ for $\alpha$-attractor E, and T model respectively. 
\section{\textbf{Dynamics of perturbative Reheating after inflation}}\label{sc3}
After the end of inflation, the inflaton field $(\phi)$ starts to oscillate around the minimum of its potential, which typically takes the form $V(\phi)\sim\phi^{2\,n}$. During this oscillatory phase, the inflaton transfers its energy to a thermal bath of SM particles, initiating the reheating process. The dynamics of the inflaton field during reheating can be expressed as \cite{Shtanov:1994ce,Ichikawa:2008ne}
\beq
\phi(t) = \phi_0(t) \mathcal{P}(t),
\eeq
where \(\phi_0(t)\) denotes the slowly varying oscillation amplitude, and \(\mathcal{P}(t)\) captures the oscillatory behavior of the inflaton field. The oscillatory function \(\mathcal{P}(t)\) can be represented via a Fourier mode expansion,  
\beq
\mathcal{P}(t) = \sum^{\infty}_{\nu=-\infty} \mathcal{P}_{\nu} \,e^{i\,\nu\,\omega\, t},
\eeq  
where $\omega$ is the oscillation frequency, and for a potential  $V(\phi)\sim\phi^{2\,n}$, this frequency is given by
\cite{Garcia:2020wiy},
\be \label{fre}
\omega = m_\phi(t)\,\gamma\,,~~~\mbox{where}~~\gamma=\sqrt{\frac{\pi\,n}{(2\,n-1)}}\frac{\Gamma\left(\frac{1}{2}+\frac{1}{2\,n}\right)}{\Gamma\left(\frac{1}{2\,n}\right)}\,.
\ee
The effective mass of the inflaton $m_\phi$ is defined by the curvature of the potential around $\phi_0$,  $m_{\phi}^2 ={\partial^2 V(\phi)}/{\partial \phi^2}|_{\phi_0}$ \cite{Garcia:2020eof}, and we have
\begin{eqnarray} \label{mphi}
m_\phi^2\simeq \frac{2n(2n-1)\Lambda^4}{\alpha_1^{2n}\, M_{\rm pl}^2}\left[\frac{\phi_0}{M_{\rm pl}}\right]^{2n-2} 
\sim {(m_\phi^{\rm end})}^2\left[\frac{a}{a_{\rm end}}\right]^{-6\,w_\phi}\,,
\end{eqnarray}
where $\aend$ and $m_\phi^{\rm end}$ denote the scale factor and the inflaton mass at the end of the inflation, respectively. The inflaton mass at the end of inflation is given by
\be\label{massend}
m_\phi^{\rm end}\simeq  \frac{\sqrt{2n\,\left(2n-1\right)}}{\alpha_1}\frac{\Lambda^{\frac{2}{n}}}{M_{\rm p}}\left(\rho_\phi^{\rm end}\right)^{\frac{n-1}{2\,n}}\,.
\ee
The average energy density of the inflaton ($\rho_{\rm\phi}$) is defined with respect to $\phi_0$ as $\rho_{\phi} = (1/2)\langle({\dot \phi}^2 + V(\phi))\rangle = V(\phi)|_{\phi_0}$. Here, the angle brackets denote averaging over many oscillations. We assume that the decay of the inflaton is sufficiently slow such that the oscillation time scale is much shorter than the time scales associated with decay and Hubble expansion. Under this adiabatic approximation, the oscillating inflaton behaves like a fluid with an effective EoS parameter $w_{\phi} = {(n-1)}/{(n+1)}$.

To constrain the microphysical parameters of reheating, one must analyze the dynamical evolution of the inflaton and radiation energy densities. This is governed by a set of coupled Boltzmann equations, supplemented by the Friedmann equation. Under the assumption of perturbative decay of the inflaton into radiation, the relevant equations are:
\begin{subequations}{\label{Boltzman}}
\begin{align}
& \dot{\rho_{\rm \phi}}+3H(1+w_\phi)\rho_\phi= -\Gamma_\phi\rho_{\rm \phi}\,(1+w_{\rm \phi}) ,  \label{Boltzman1} \\
&\dot{\rho}_{\rm R}+4H\rho_{\rm R}=\Gamma_{\rm \phi}\rho_{\rm \phi}(1+w_{\rm \phi})\,,
\label{Boltzman2} \\
&H^2=\frac{\rho_{\rm \phi}+\rho_{\rm R}}{3\,M_{\rm pl}^2}\label{Boltzman3}\,,
\end{align}
\end{subequations}
where, $\Gamma_{\phi}$ is the perturbative inflaton decay rate. During reheating, inflation is decaying into radiation, and for our study, we consider radiation to be massless SM bosonic particles ($b$). To model the inflaton decay, we consider two representative interaction processes, described by the following phenomenological interaction Lagrangian
\be
\mathcal{L}_{\rm int} \supset  \begin{cases} 
 g\, \phi\, b^2\,, & \phi \to b\, b\,,\\
 \mu\, \phi^2\, b^2\,, & \phi\,\phi \to b\, b\,.\\
\end{cases}
\label{RH:procs}
\ee
\begin{figure}[t]
\centering
\includegraphics[width=\columnwidth]{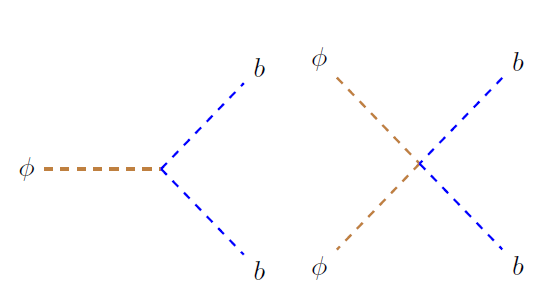}
\caption{\em \small Feynman diagrams for the inflaton decay and annihilation channels:(i) $\phi\rightarrow\,b\,b$, (ii) $\phi\,\phi\rightarrow b\,b$ }
\label{Feynmann}
\end{figure}
\begin{table*}[t]
\scriptsize{
\caption{Numerical values of the Fourier sums in the effective couplings:}\label{DMfouriersum}
\centering
 \begin{tabular}{||c | c |c |c |c| c| c||} 
 \hline
 $n(w_\phi)$ & $ \sum \nu \lvert\mathcal P_\nu \rvert^2 $ & $ \sum \nu\lvert\mathcal P^{(2)}_\nu \rvert^2 $ & $ \sum \lvert\mathcal P_\nu \rvert^2 $ &  $ \sum \lvert\mathcal P^{(2)}_\nu \rvert^2 $& $ \sum \lvert\mathcal P^{(2n)}_\nu \rvert^2 $ & $ \sum \frac{1}{\nu^2}\lvert\mathcal P^{(2n)}_\nu \rvert^2 $  \\ [0.5ex] 
 \hline\hline
 1 (0.0) & 0.25 & 0.125 & 0.25 &0.0625  & 0.0625 & 0.0156\\ \hline
 2 (1/3) & 0.229 & 0.1256 &0.2284  &0.0623  & 0.0635 & 0.0145\\ \hline
 3 (0.5) & 0.218  & 0.124 & 0.2155 & 0.061  & 0.056  & 0.011 \\ \hline
 6 (5/7)&0.2  & 0.118 &0.1966  & 0.056  & 0.038   & 0.00534 \\  \hline
 15 (7/8) & 0.186 & 0.11 & 0.181 &0.051  & 0.01906   & 0.00132\\ [1ex] 
 \hline
 \hline
 \end{tabular}}
\end{table*}
Here, $\mu$ is a dimensionless quartic coupling constant, while $g$ is a dimensionful trilinear coupling with mass dimension one, characterizing the strength of the respective interaction processes. The associated Feynman diagrams are depicted in Fig.\ref{Feynmann}.
Incorporating the Bose enhancement (BE) effect due to the bosonic nature of the decay products, the effective decay width of the inflaton can be written as \cite{Drewes:2019rxn,Garcia:2020wiy,Haque:2023yra}
\be
\label{Eq:gammaphibis}
\Gamma_{\phi} = 
\begin{cases}
\frac{g^2\,(2\,n+2)(2\,n-1)\,\gamma}{8\,\pi\,m_\phi(t)}\sum \nu\,\lvert\mathcal{P}_\nu\rvert^2\left(1+2\,f_\nu(t)\right),\\
\frac{\mu^2\,2n\,(2\,n+2)(2\,n-1)^2\,\gamma\,\rho_\phi(t)}{8\,\pi\,m^3_\phi(t)}\sum \nu\lvert\mathcal{P}^2_\nu\rvert^2\left(1+2\,f_\nu(t)\right),
\end{cases}
\ee
where $f_\nu(t)=\left(e^{\frac{\nu\,\gamma\,m_\phi(t)}{2\,T}}-1\right)^{-1}$ is the equilibrium Bose-Einstein distribution function. This enhancement can significantly impact the thermal history during reheating, making the reheating process more efficient compared to scenarios where the Bose-Einstein statistical effects are neglected (see Refs.~\cite{Haque:2023yra,Drewes:2017fmn,Adshead:2019uwj}).  From the expression, it is clear that when the radiation temperature exceeds the inflaton mass, i.e., ($ T_{rad}\gg m_\phi(t)$), the BE effect becomes significant. Under this assumption, the inflaton decay width can be approximated as
\begin{subequations}{\label{decaywidth}}
\begin{align}
&\Gamma_{\phi\rightarrow b\,b}=\frac{g^2_{\rm eff}}{8\,\pi\, m_\phi(t)}\frac{4\,T}{m_\phi(t)}\,\\
 & \Gamma_{\phi\,\phi\rightarrow b\,b}=\frac{\mu^2_{\rm eff}\,\rho_\phi(t)}{8\,\pi\, m^3_\phi(t)}\frac{4\,T}{m_\phi(t)}\,
\end{align}
\end{subequations}
where
\begin{subequations}{\label{t2}}
\begin{align}
&\left(\frac{g_{\rm eff}}{g}\right)^2=(2\,n+2)(2\,n-1)\,\sum^{\infty}_{\nu=1}\lvert\mathcal P_\nu\rvert^2\\
&\left(\frac{\mu_{\rm eff}}{\mu}\right)^2=2\,n(2\,n+2)(2\,n-1)^2\sum^{\infty}_{\nu=1}\lvert\mathcal{P}^{(2)}_\nu\rvert^2.
\end{align}
\end{subequations}
Here, the Fourier coefficients \( \mathcal{P}_\nu^{(2)} \) are defined through the harmonic expansion of the squared inflaton field,
\begin{eqnarray} \phi^2(t) &=& \phi_0^2(t)\,\mathcal{P}^2(t) = \phi_0^2(t)\sum_{\nu=-\infty}^{\infty} \mathcal{P}^{(2)}_\nu e^{i\,\nu\,\omega t}. 
\end{eqnarray}
For the analytical treatment, we incorporate the decay width (Eqs.~\ref{decaywidth}) into the Boltzmann equations given in Eq.~\eqref{Boltzman}, and solve for the evolution of the energy densities of the inflaton and radiation components as
\begin{eqnarray}\label{rhophi}
&&\rho_\phi \simeq 
\rho^{end}_{\phi}A^{-3(1+w_{\phi})} ,~~\mbox{where}~~A=a/a_{end}\,,\\
&&\rho_{\rm R} \simeq
\begin{cases}
\left[\frac{3g^2_{\rm eff}M^2_p(1+w_\phi)H_{end}}{4\pi\epsilon^{1/4}(1+3w_\phi)(m^{end}_\phi)^2A^3}\left(A^{\frac{3(1+3w_\phi)}{2}}-1\right)\right]^{4/3}~\,\nonumber\\
\left[\frac{9\mu^2_{\rm eff}M^4_p(1+w_\phi)H^3_{end}}{8\pi\epsilon^{1/4}(5w_\phi-1)(m^{end}_\phi)^4A^3}\left(A^{\frac{3(-1+5w_\phi)}{2}}-1\right)\right]^{4/3} .
\end{cases}
\end{eqnarray}
With these solutions, we can identify important physical quantities, namely, the reheating temperature $T_{\rm re}$, which is defined at the end of the reheating phase when the inflaton and radiation energy densities become equal, $\rho_{\phi}(\Are) = \rho_R(\Are)=(\pi^2/30)g^{\rm re}_\star \tre^4=\epsilon\,\tre^4$. This condition marks the transition from an inflaton-dominated phase to a radiation-dominated era that sets the stage for the standard hot Big Bang nucleosynthesis (BBN). Now, let us define the reheating temperature in this context. For the bosonic decay process $\phi\rightarrow b\,b$, the reheating temperature $\tre$ is defined as
 \begin{equation}{\label{bedecay}}
\begin{aligned}
&T_{re}=\left[\frac{3M^2_p(1+w_\phi)H_{end}\,g^2_{\rm eff}}{4\pi\epsilon(1+3w_\phi)(m^{end}_\phi)^2} A_{re}^{-\frac{3}{2}(1-3w_\phi)}\right]^{1/3}\,\\
&A_{re}=\left(\frac{4\pi(1+3w_\phi)(m^{end}_\phi)^2}{(1+w_\phi)\,g^2_{\rm eff}(\frac{\epsilon}{3})^{1/4}}\left(\frac{H_{end}}{M_p}\right)^{1/2}\right)^{\frac{4}{3(1+9w_\phi)}}\,.
\end{aligned}
\end{equation}
Whereas for bosonic annihilation \footnote{The expressions for $\tre$ defined in Eq.~\ref{beannihi} is not valid for $\wphi<3/13$, as reheating is not possible when $\wphi<3/13$ \cite{Haque:2023yra}. Similarly, in the absence BE effect, reheating is also not possible when $\wphi<1/9$\cite{Garcia:2021gsy,Haque:2023yra}}, $\phi\,\phi\rightarrow b\,b$
\begin{equation}{\label{beannihi}}
\begin{aligned}
&T_{re}=\left[\frac{9M^4_p(1+w_\phi)H^3_{end}\,\mu^2_{\rm eff}}{8\pi\epsilon(5w_\phi-1)(m^{end}_\phi)^4}A^{\frac{-3(3-5w_\phi)}{2}}_{re}\right]^{1/3}\,\\
&A_{re}=\left(\frac{8\pi(27\epsilon)^{1/4}(5w_\phi-1)}{9(1+w_\phi)M^{5/2}_p\,H^{3/2}_{end}\mu_{\rm eff}^2}(m^{end}_\phi)^4\right)^{\frac{4}{3(13w_\phi-3)}}\,.
\end{aligned}
\end{equation}
For a higher inflaton EoS, specifically when $\wphi\geq1/3$, the BE effect remains effective throughout the entire range of reheating temperatures- from $\tmax$ to $\tbbn$ (see for details  \cite{Haque:2023awl}). Therefore, the derived analytical expression for $\tre$, which incorporates the BE effect, can be safely used to find the $\tre$ when $\wphi\geq1/3$. However when $\wphi<1/3$, the condition $T_{\rm rad}>\mphi$ does not hold over the entire range of $\tre$, especially at lower $\tre$ . In such cases, the BE effect becomes negligible, and one must revert to the zero-temperature approximation for the inflaton decay width. Under this assumption, the analytical expression for the $\tre$,
\begin{eqnarray}{\label{wbedecay}}
&&T_{re}=\left\{\begin{array}{ll}
\left(\frac{6M^2_p(1+w_\phi)H_{end}}{8\pi\epsilon(5+3w_\phi) m^{end}_\phi}g^2_{\rm eff,0}\right)^{1/4}A^{-\frac{3}{8}(1-w_\phi)}_{re}~~\phi\rightarrow b\,b\,,\\
\left(\frac{9 M^4_p(1+w_\phi)H^3_{end}}{4\pi\epsilon(9w_\phi-1)( m^{end}_\phi)^3}\mu^2_{\rm eff,0}\right)^{1/4} A^{\frac{9(w_\phi-1)}{8}}_{re}~~\phi\,\phi\rightarrow b\,b\,,\\
\end{array}\right.
\end{eqnarray} 
where
\begin{eqnarray}{\label{wbeannihi}}
&&A_{re}=\left\{\begin{array}{ll}
&\left(\frac{4\pi(5+3w_\phi)H_{end}m^{end}_\phi}{(1+w_\phi)g^2_{\rm eff,0}}\right)^{\frac{2}{3+9w_\phi}}~~\phi\rightarrow b\,b\,,\\
&\left(\frac{4\pi(9w_\phi-1)(m^{end}_\phi)^3}{3(1+w_\phi)M^2_pH_{end}\mu^2_{\rm eff,0}}\right)^{\frac{2}{3(5w_\phi-1)}}~~\phi\rightarrow b\,b\,,
\end{array}\right.
\end{eqnarray}\\
\begin{eqnarray}{\label{t2}}
&&\left(\frac{g_{\rm eff,0}}{g}\right)^2=(2n+2)(2n-1)\,\gamma\sum^{\infty}_{\nu=1}\nu\lvert\mathcal P_\nu\rvert^2\nonumber\\
&&\left(\frac{\mu_{\rm eff,0}}{\mu}\right)^2=2n(2n+2)(2n-1)^2\gamma\sum^{\infty}_{\nu=1}\nu\lvert\mathcal P^{(2)}_\nu\rvert^2.
\end{eqnarray}

We have taken the effective number of relativistic degrees of freedom at the end of reheating, $g_*^{\rm re} =427/4$ throughout the paper. As mentioned earlier in the introduction, there exists a naive lower limit on the temperature $T^{\rm min}_{\rm re} = T_{\rm BBN} = 4\, {\rm MeV}$ \cite{Kawasaki:2000en,Hannestad:2004px}

The post-reheating history plays a crucial role in constraining both inflationary and post-inflationary (reheating) parameters. A standard assumption in cosmology is that the comoving entropy density remains conserved from the end of reheating to the present epoch. This conservation leads to an essential relation among the parameters ($N_{\rm k},\,N_{\rm re},\,T_{\rm re}$) as follows \cite{Dai:2014jja,Cook:2015vqa}, 
\be \label{entropy-conservation}
T_{\rm re}=\left(\frac{43}{11\,g_*^{\rm re}}\right)^{1/3}\,\left(\frac{a_0\,H_{ k}}{k}\right)\,e^{-(N_k+N_{\rm re})}\,T_0,
\ee
where the present CMB temperature is $T_0=2.725$ K. $N_{\rm re}=\ln (A_{\rm re})$ is the e-folding number for reheating and $N_{\rm re}=0$ corresponds to an instantaneous reheating scenario ($\tre\simeq10^{15}$ GeV). The relation (Eq.~\ref{entropy-conservation}) serves as a powerful tool to constrain the reheating temperature ($\tre$) by connecting them to the inflationary parameters ($N_{\rm k}\,,H_k$), thereby enabling one to constrain them using the CMB observables ($\ns\,,r$). Furthermore, we can constrain the inflaton coupling parameters ($g,,\mu$) by connecting Eqs.~\ref{entropy-conservation} with the reheating temperature expressions defined for  decay (Eqs.~\ref{bedecay},\,\ref{beannihi}) and annihilation processes (Eqs.~\ref{wbedecay},\,\ref{wbeannihi}).

So far, we have discussed the connection between inflationary and reheating parameters through large-scale inflationary observables $\ns$ and $r$. Interestingly, the inflationary energy scale and the reheating temperature can, in principle, also be directly probed through observations of the PGWs background. In the following section, we demonstrate how constraints on extra relativistic degrees of freedom at the time of BBN, arising from PGWs, can further restrict the model parameters.

\section{PGWs and $\mathbf{\Delta N_{\rm eff}}$ constraints}\label{sc4}
Primordial gravitational waves (PGWs) are essential cosmic relics that offer direct insight into the unknown physics of the early universe. They can originate from quantum fluctuations during inflation \cite{Grishchuk:1974ny,Guzzetti:2016mkm,Starobinsky:1979ty}. Owing to their extremely weak interactions, PGWs propagate through the universe almost unimpeded and hence act
as a unique probe of the very early universe, such as the inflation and post-inflationary reheating phase \cite{Mishra:2021wkm,Benetti:2021uea,Haque:2021dha,Maity:2024cpq}.  The amplitude and the evolution of the PGWs spectrum are sensitive to the energy scale of the inflation and the inflaton EoS $w_\phi$.

For those modes between $k_{\rm re} < k < k_{\rm end}$ which become sub-Hubble at some time during reheating, the PGW spectrum at the present time assumes the following form, 
(see \cite{Haque:2021dha} for detailed calculation)
\be \label{GW}
\Omega^{k}_{\rm GW}h^2\simeq \Omega^{\rm inf}_{\rm GW}h^2 \frac{\mu (w_\phi)}{\pi}\left(\frac{k}{ k_{\rm re}}\right)^{-\frac{(2-6\,w_\phi)}{(1+3\,w_\phi)}}
\ee
Where, $\mu(w_\phi)=(1+3\omega_\phi)^{\frac{4}{1+3\,w_\phi}}\,\Gamma^2\left(\frac{5+ 3\,w_{\phi}}{2+6\,w_{\phi}} \right)$, which typically ${\cal O}(1)$ value.
The scale-invariant part $\Omega^{\rm inf}_{\rm GW}h^2$,
is fixed  by the inflationary energy scale as follows, 
\be \label{omegainf}
\Omega^{\rm inf}_{\rm GW}h^2
= \frac{\Omega_{\rm R} h^2 H_{\rm end}^2}{12 \pi^2 M_{\rm p}^2}\, = 6 \times 10^{-18} \left(\frac{H_{\rm end}}{10^{13}\,\mbox{GeV}}\right)^2.
\ee
Where we used the present radiation abundance $\Omega_{\rm R} h^2= 4.16\times10^{-5}$. One can observe that the spectrum is blue (red) tilted for $w>1/3\,(<1/3)$.
However, regardless of its origin, the energy density of PGWs acts as radiation at the time of BBN, and thus its impact on BBN is fully captured in terms $\Delta N_{\rm eff}$. From the the recent ACT observation the $\Delta N_{\rm eff}$ is tightly constrained, $\Delta N_{\rm eff}\leq0.17$ at $2\sigma$ \cite{ACT:2025tim,ACT:2025fju}. This is where the BBN bound related to the effective number of relativistic degrees of freedom comes into play, and this translates into the bound on PGW spectrum as $\Omega^{k_{\rm end}}_{\rm GW}\,h^2\leq9.5\times10^{-7}$ \cite{Pagano:2015hma,Yeh:2022heq,Planck:2018vyg}.

The constraint equation on PGW spectrum in terms of $\Delta N_{\rm eff}$ can be expressed as,
\begin{align}
\Delta N_{\rm eff} \geq \frac {1}{\Omega_{\rm R}h^2} \frac{8}{7}\left(\frac{11}{4}\right)^{4/3} \int_{k_{\rm 0}}^{k_{\rm f}}\frac{dk}{k}\Omega_{\rm GW}^{\rm k}\,h^2(k)
\end{align}
Using the PGWs spectrum expression in the above expression, the above
condition reduces to,
\be \label{bbncon}
2.1\times 10^{-11} \frac{\mu (w_\phi) (1+3\,w_\phi)} {\pi(6\,w_\phi -2)}  \left[\frac{H_{\rm end}}{10^{-5} M_{\rm p}}\right]^2 \leq \left(\frac{k_{\rm end}}{ k_{\rm re}}\right)^{\frac{(2-6\,w_\phi)}{(1+3\,w_\phi)}}
\ee
The relation between the two scales $(k_{\rm end}, k_{\rm re}$) can be further expressed in terms of reheating temperature as follows: 
\be
\left(\frac{k_{\rm f}}{k_{\rm re}}\right)=\left(\frac{15\,\rho_\phi^{\rm end}}{\pi^2\,g_*^{\rm re}}\right)^{\frac{1+3 w_\phi}{6\,(1+w_\phi)}}\,T_{\rm re}^{\frac{-2 (1+3 w_\phi)}{3(1+w_\phi)}}\,.
\ee
 Finally, using the above relations into Eq.\ref{bbncon}, we obtain a lower limit on the reheating temperature, particularly when the inflaton equation of state $w_{\phi} >1/3$, 
\begin{eqnarray}{\label{tregw1}}
T_{\rm re}&>&0.35 \left(\frac{45.6\,M_p^4}{\mu(\phi)}\right)^{\frac{3\,(1+w_\phi)}{4(1-3\,w_\phi)}}\left(\rho_\phi^{\rm end}\right)^{-\frac{1}{2}\,\frac{(1+3\,w_\phi)}{(1-3\,w_\phi)}}\nonumber\\
&=& T_{\rm re}^{\rm GW}.
\end{eqnarray}
By setting the above temperature equal to the BBN energy scale, $T_{\rm re}^{\rm GW} \sim T_{\rm BBN} \sim 4$ MeV, we find that the BBN bound on PGWs becomes relevant only when $w_\phi \geq 0.60$. We denote this new lower bound on the reheating temperature, arising from the PGW constraint, as $T_{\rm re}^{\rm GW}$.

Since $T_{\rm re}^{\rm GW}$ depends on both $H_{\rm end}$ and $w_\phi$, for a given of $w_\phi$ it can restrict inflationary energy scale, which in turn puts a constraint on the potential parameter such as $\alpha$  for the attractor model. Another interesting point is that this $T_{\rm re}^{\rm GW}$ also set bounds on the inflaton coupling to SM fields, which we will discuss in more detail in the subsequent sections for inflaton decay ($\phi\to b\,b$) or annihilation ($\phi\,\phi\to b\,b$)processes.
\section{Nonperturbative constraints  during reheating}\label{sc6}
Radiation can be resonantly produced during reheating if the inflaton–SM coupling is sufficiently large, leading to non-perturbative particle production by parametric resonance effect. However, the present analysis so far is restricted to the perturbative regime. Therefore, it is crucial to identify the coupling range—denoted as the non-perturbative constraint (NPC)—beyond which our perturbative treatment may no longer be valid. To determine this boundary, we shall closely follow the discussion from \cite{Chakraborty:2023ocr, Chakraborty:2023lpr}, where we investigate the dynamics of the radiation field fluctuations in the background of an oscillating inflaton condensate. This leads to a mode equation for the radiation field that takes the form of a generalized Hill-type differential equation:
\begin{align}\label{fluctueq}
\frac{1}{(m_{\phi}^{\rm end})^2}\frac{d^2 X_{\rm k}}{dt^2}+{\cal Q}^2(k,t) X_{\rm k}=0
\end{align}
Where, $X_{\rm k}$ is a particular field mode for  boson $(b)$, and the associated dimensionless time-dependent frequencies are
\be \label{resop}
{\cal Q}^2(k,t) = \begin{cases} 
\frac{k^2}{(m_{\phi}^{\rm end})^2 a^2}+ q^2(t) \mathcal{P}(t) &~~\phi\rightarrow b\, b \\
\frac{k^2}{(m_{\phi}^{\rm end})^2 a^2}+ q^2(t) \mathcal{P}^2(t)
& ~~\phi\,\phi \rightarrow{b\,b} \\
\end{cases}
\ee
Given the inflaton-radiation coupling of our interest Eq.\ref{RH:procs}, the resonance parameter $q$ is identified as 
\be\label{qt}
q(t) = \begin{cases} 
\sqrt{\frac{g \phi_0(t)}{(m_{\phi}^{\rm end})^2}} & ~~~~\phi\rightarrow b\,b\\
 \sqrt{\frac{\mu \phi^2_0(t)}{(m_{\phi}^{\rm end})^2}} & ~~~~\phi\,\phi\rightarrow b\,b\\
\end{cases}
\ee
\begin{figure*}[t]  
\includegraphics[width=18.0cm,height=7.5cm]{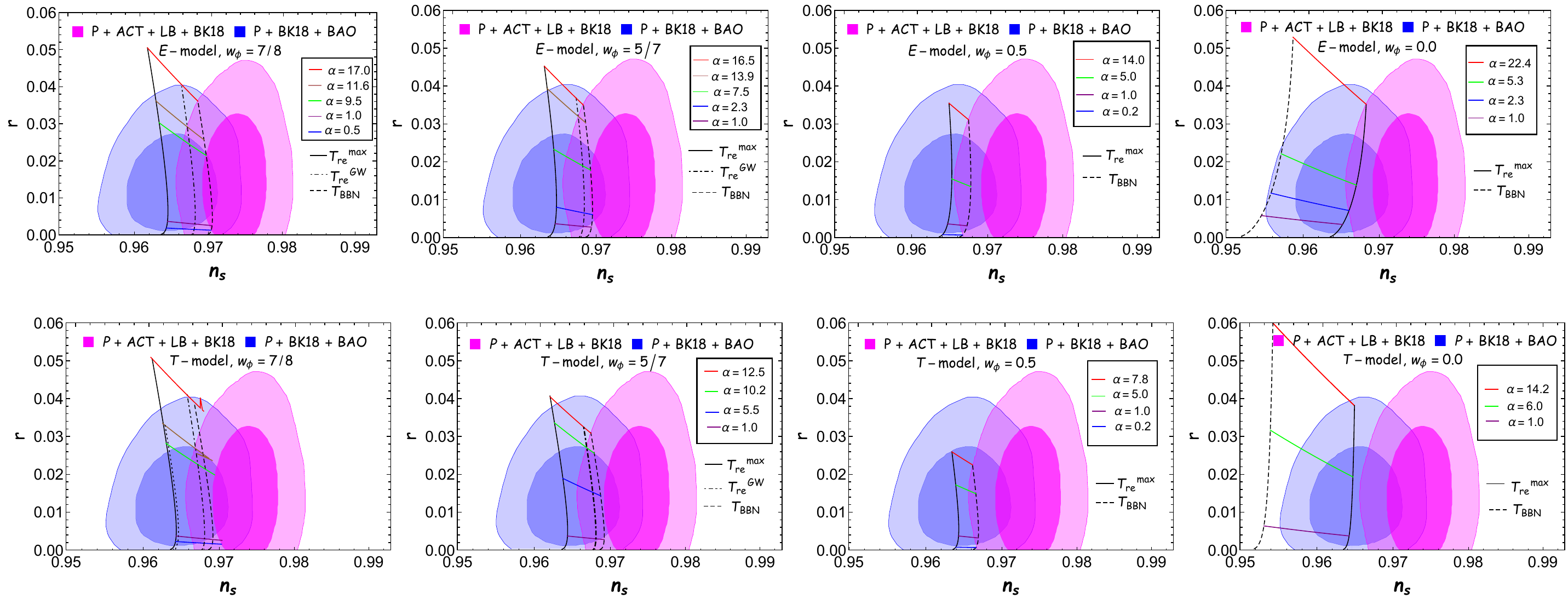}
   \caption{\em $\alpha$-attractor model predictions of $(n_s, r)$ for various combinations of $(\alpha,\,w_\phi)$ in the $(\ns - r)$ plane, and compared with recent combined observational data from {\rm 
\textbf{P+ACT+LB+BK18}} (magenta), while the blue region represents the older constraints from Planck \cite{Tristram:2021tvh}. These predictions span reheating temperatures ranging from the maximum possible reheating temperature $T_{\rm re}^{\rm max} \sim 10^{15}$ GeV (solid black lines) to the minimum possible reheating temperature $T_{\rm re}^{\rm min}=\tbbn \sim 4$ MeV (dashed black lines). The deep magenta/blue and light magenta/blue shaded regions represent the $1\sigma$ ($68\%$ CL) and $2\sigma$ ($95\%$ CL) confidence intervals, respectively. Overproduction of GW at the time of BBN provides a constraint on reheating temperature, defined as  $T_{\rm re}^{\rm GW}$, which is shown in a dot-dashed black line.}
\label{nsrE}
\end{figure*}
In the literature, conditions of resonance are usually derived in the Minkowski background \cite{Kofman:1997yn,Greene:1997fu,Greene:2000ew}, and the resonance is broadly classified into $q>1$ called the broad resonance regime, and $q \lesssim 1$ called the narrow or weak resonance regime. However, in the real scenario, the resonance parameter $q$ depends non-trivially on time through inflaton oscillation amplitude $\phi_0(t)$, and hence the non-constancy of the resonance strength defining parameter $q$ makes the naive Minkowski approximation insufficient to get a strong resonance effect in a dynamical background. Particularly, with increasing reheating equation of state, the inflaton amplitude dilutes very fast, $\phi_0(t)\propto \phi_{\text{end}} ({a(t)}/{a_{\text{end}}})^{-{6}/{(n+2)}}$ depending on different $n$ values. Using decaying inflaton amplitude in $H^2\simeq {V(\phi_0)}/{3M_{p}^2}$, we get the leading order behavior of the post-inflationary scale factor as
\begin{equation}\label{scaleforn}
  a(t)=  a_{\text{end}} \left(\frac{ t}{ t_{\rm end}}\right)^{\frac{n+1}{3n}} .
\end{equation}
Using above equation and Eq.\ref{qt} the resonance parameter $q(t)$ evolves as,
\be\label{qtA}
q(t) = \begin{cases} \sqrt{\frac{g \phi_{\text{end}}}{(m_{\phi}^{\rm end})^2}}\left(\frac{t}{t_{\rm end}}\right)^{-\frac{1}{2\,n}} &~~{\rm for}~~\phi\rightarrow b b \\
 \sqrt{\frac{\mu \phi^2_{\text{end}}}{(m_{\phi}^{\rm end})^2}}\left(\frac{t}{t_{\rm end}}\right)^{-\frac{1}{n}} & ~~{\rm for}~~\phi\phi\rightarrow bb \\
\end{cases}
\ee
\begin{figure}[t]
\centering
\includegraphics[width=6.0cm, height=4cm]{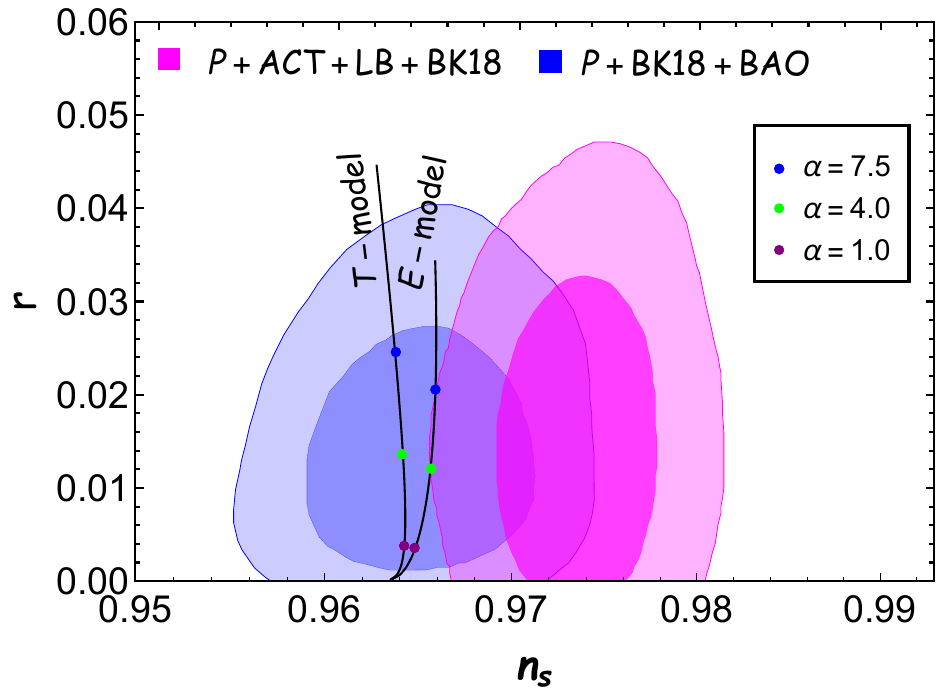}
\caption{\em $\alpha$-attractor model predictions of $(n_s, r)$ for $\wphi=1/3$ in the $(\ns - r)$ plane, and compared with recent combined observational data from {\rm 
   \textbf{P+ACT+LB+BK18}}, while the blue region represents the older
constraints from Planck.}
\label{w13}
\end{figure}
Resonant particle production occurs when the adiabaticity condition is violated($|\frac{\dot{\mathcal{Q}}(k,t)}{\mathcal{Q}^2(k,t)}|>>1$), typically during inflaton zero-crossings in its oscillatory phase. Two key conditions must be fulfilled to achieve efficient resonance in a dynamical background: (i) the inflaton must undergo multiple oscillations within the relevant timescale, and (ii) the resonance parameter $q$ must remain larger than unity during that period. Broad resonance is realized if $q$ evolves from its initial large value to unity over at least one full oscillation. Using this dynamical criterion, we derive a lower bound on the inflaton–radiation coupling for a general background equation of state. To estimate the number of oscillations required for $q$ to decrease from initial value $q_{\text{in}}$ to 1, we define the dimensionless time period of inflaton oscillation as $T^{(\Omega)} = 2\pi m_{\phi}^{\text{end}} / \Omega_0$, where $\Omega_0$ is the oscillation frequency evaluated at $\phi_0 = M_p$ (see Eqs.~\ref{fre} and \ref{mphi}). The number of oscillations is then given by,
\begin{align}\label{noosci}
     N_{\text{osc}}= \frac{t -t_{\rm end}}{T^{(\Omega)}} =
   \begin{cases}
  & \frac{t_{\rm end}}{T^{(\Omega)}}\left(\left(\frac{\sqrt{\mu}\,\phi_{\text{end}}}{m_{\phi}^{\text{end}}}\right)^{n}-1\right)\\
   &\frac{t_{\rm end}}{T^{(\Omega)}}\left(\left(\frac{\sqrt{g\,\phi_{\text{end}}}}{m_{\phi}^{\text{end}}}\right)^{2\,n}-1\right) ,
   \end{cases}
\end{align}
Where
\[\frac{t_{\rm end}}{T^{(\Omega)}}=\frac{\sqrt{2n(2n-1)}(n+1)M_{\rm p} \Omega_{0}} {2\pi\sqrt{3} \,n\,m_{\phi}^{\text{end}}\,\phi_{\text{end}}}.\] 
Therefore, the minimum criterion that has to be met to achieve efficient resonance is $N_{\text{osc}}>1$. This yields the lower bound of the coupling strength for two production channels as
\begin{eqnarray}\label{lowerboundgh}
& \tilde{g}=g/\mphi^{\rm end} > \frac{m_{\phi}^{\rm end}}{\phi_{\text{end}} }\bigg[1+\frac{2\pi\sqrt{3}  \,n\,\phi_{\text{end}}\,m_{\phi}^{\rm end}}{\sqrt{2n(2n-1)}(n+1)M_{\rm p} \Omega_{0}}\bigg]^{\frac{1}{n}}\nonumber\\
 &\mu > \left(\frac{m_{\phi}^{\rm end}}{\phi_{\text{end}} }\right)^2 \bigg[1+\frac{2\pi\sqrt{3} \,n\,\phi_{\text{end}}\,m_{\phi}^{\rm end}}{\sqrt{2n(2n-1)}(n+1)\,M_{\rm p} \Omega_0}\bigg]^{\frac{2}{n}}\,\label{nonper2}
\end{eqnarray}
In the above equation\,\ref{lowerboundgh}, the second term inside the bracket appears as a correction term in an expanding inflaton background, which depends upon the background oscillation frequency, and the background EoS in the post-inflationary phase. This correction term becomes important for higher post-inflationary EoS. In the context of the attractor model, the typical upper bounds \footnote{If the coupling parameter is sufficiently large, one might expect it to modify the tree-level inflationary potential through one-loop Coleman-Weinberg (CW) corrections \cite{PhysRevD.7.1888,Drees:2021wgd}. However, earlier work \cite{Chakraborty:2023ocr} has shown that, in the case of bosonic decay, such corrections do not significantly alter the tree-level potential. We have checked that the same conclusion applies to our analysis for bosonic annihilation.} on the dimensionless coupling parameters are $\tilde{g}\sim (10^{-4}, 10^{-3}) $, and $\mu\sim (10^{-9},\, 10^{-6})$ with the range of $w_\phi=(0-1)$. For both channels, NPC imposes a stronger bound on the coupling parameter than the coupling parameter associated with the instantaneous reheating temperature $\tre\simeq10^{15}$ GeV (see Fig.~\ref{trilinear},\ref{quartic}).

One of the main motivations of this work is to probe the inflaton–SM coupling using current ACT observations. To extract the coupling parameters, we use the relation between the reheating temperature $(T_{\rm re})$ and the inflaton coupling obtained within the perturbative reheating framework. In the non-perturbative regime, however, this relation can be significantly modified due to resonant particle production. Therefore, when the reheating temperatures allowed by ACT observations lie within the non-perturbative (NPC) region, the perturbative relation cannot be used to reliably determine the coupling parameters. A proper study of this regime would require a dedicated lattice analysis to capture the non-linear dynamics of reheating. Hence, we defer the investigation of these non-perturbative effects to future work.

    \begin{table*}[ht]
\centering
\caption{Allowed ranges of the inflationary parameters ($\alpha$, $H_{\rm end}$, $N_{\rm k}$) and the reheating temperature ($T_{\rm re}$) for the $\alpha$-attractor model, constrained by combined P+ACT+LB+BK18 observations along with BBN and PGW bounds.}
\label{alphatable}
\begin{tabular}{|c|c|c|c|c|c|c|}
\hline
Model & $n (w_\phi)$ & $\alpha$ & $H_{\text{end}}\times 10^{12} \,\text{(GeV})$ & $N_k$ & $T_{\text{re}}\,\text{(GeV})$ \\ \hline
\multirow{5}{*}{E} & 1 (0) & $2.2-22.5$ & $22.1-47.5 $& $53.93-56.36$  & $2.6\times10^{11}-10^{15}$  \\ \cline{2-6}
& 2 (1/3) &$4.0-7.5$ & $8.42-8.62$ &$53.12-53.89$  & $4.0\times10^{-3}-10^{15}$  \\ \cline{2-6}
 & 3 (1/2) &$ 0.2-13.9$ & $5.25-5.81$ & $57.18-61.56$ & $4\times 10^{-3}-4.82\times10^{12}$ \\ \cline{2-6}
& 6 (5/7) & $0-13.9$ & $2.61- 6.29$ & $58.07-61.93$ & $14.88 -1.64\times10^{12}$\\ \cline{2-6}
& 15 (7/8)  & $0-11.6$ & $2.61 - 5.59$ & $58.81-61.91$ & $1.99\times 10^{4}  -8.85\times10^{11}$ \\ \hline
\multirow{5}{*}{T} & 1 (0) & null & null & null  & null  \\ \cline{2-6}
& 2 (1/3) & null & null & null &  null   \\ \cline{2-6}
& 3 (1/2)  & $0.2-7.8$ & $4.59 - 5.81$ & $59.08-61.43$ & $4\times 10^{-3}$-$2.16\times10^5$ \\ \cline{2-6}
& 6 (5/7) & $0-10.2$ & $2.61 - 5.48$ & $59.17-61.92$ & $10.72-1.75\times10^{10}$ \\ \cline{2-6}
& 15 (7/8)  & $0-10.1$ & $2.61-5.37$ & $59.22-61.91$ & $1.99-1.51\times10^{11}$ \\ \hline
\end{tabular}
\end{table*}

\section{Constraining inflationary and post-inflationary parameters}\label{sc7}
In this section, we present the consolidated constraints on both inflationary and post-inflationary reheating parameters by incorporating observational as well as theoretical bounds. As shown in Eq.~\ref{entropy-conservation}, the CMB observables $(\ns\,,r)$) are connected to the inflationary parameter $N_k$ through the reheating temperature $\tre$. The combined observational data P+ACT+LB+BK18 place stringent constraints on large-scale inflationary observables ($\ns\,,r$), thereby constraining the inflationary energy scale and the reheating temperature. To avoid the spoilage
of BBN, the reheating temperature must also satisfy $\tre\geq T_{\rm BBN}$. In addition, we include the BBN constraint from PGWs, which sets a lower bound on $\tre$ (defined in Eq.~\ref{tregw1}). Altogether, the final constraints on the model parameters arise from a combination of \textbf{P+ACT+LB+BK18+BBN+PGWs}. Moreover, since reheating temperature is directly related to the inflaton-SM coupling parameters, the aforementioned bounds on the reheating temperature also impose indirect bounds on the inflaton decay or annihilation couplings.

 \begin{figure*}[t]
         \begin{center}
      \includegraphics[width=0017.50cm,height=7cm]{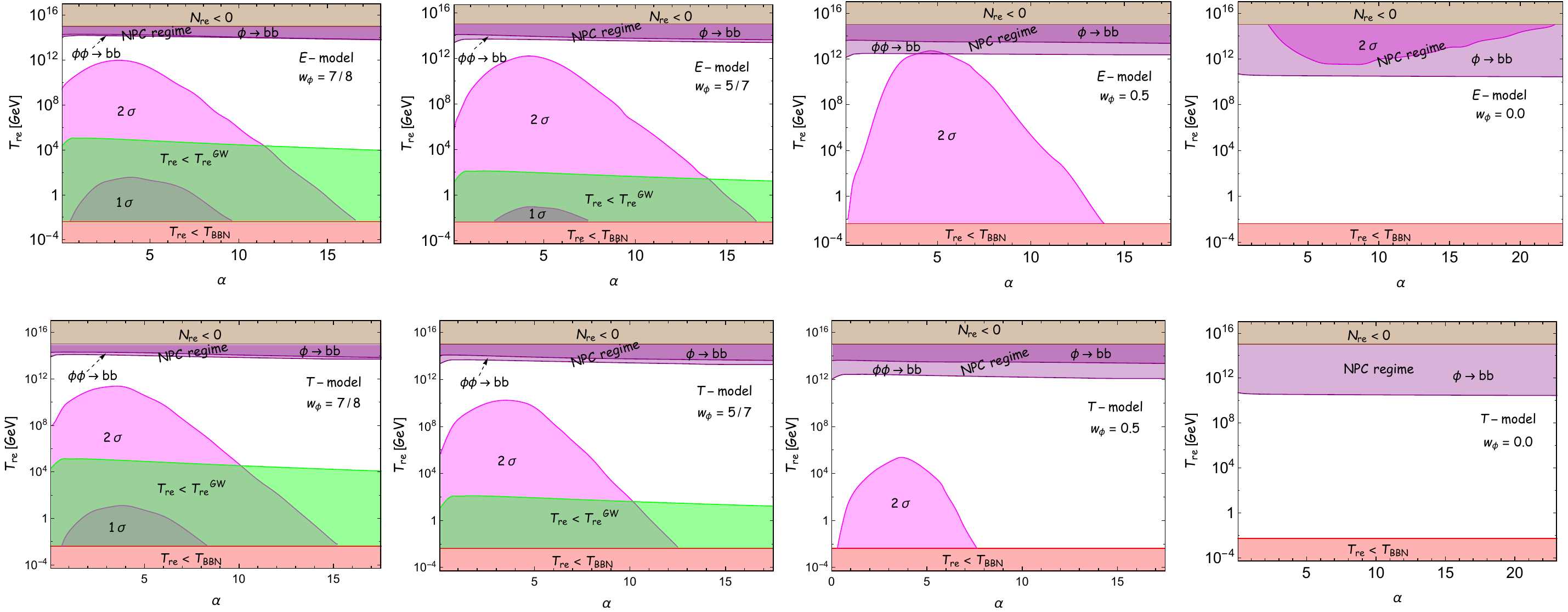}
          \caption{\em The impact of {\rm \textbf{P+ACT+LB+BK18+BBN+PGWs+NPC}} on reheating temperature $T_{\rm re}$ with respect to $\alpha$ both for  E-model (upper pannel) and T-model (lower pannel). We consider four representative values of the inflaton EoS $w_{\phi} = (7/8,5/7,0.5,0.0)$. The red and green shaded region indicates $T_{\rm re} < T_{\rm BBN}$ and $T_{\rm re} < T_{\rm re}^{\rm GW}$, respectively, respectively, indicating regions excluded by BBN and GW overproduction. The purple shaded regions depict NPC regions, the reheating temperature predicted by ACT observations within this region is not excluded; however, the corresponding inflaton coupling predictions obtained using the perturbative framework are not reliable, since the perturbative description breaks down in this regime. The brown-shaded region implies no reheating ($N_{\rm re}<0$).}
          \label{Tree}
          \end{center}
      \end{figure*}
      \begin{table*}[ht]\label{coupling table}
\centering
\caption{Allowed ranges of perturbative inflaton-SM coupling parameters ($\tilde{g}\,,\mu$) for the $\alpha$-attractor model, constrained by combined P+ACT+LB+BK18 observations along with BBN and PGW bounds.}
\label{couplingtable}

\begin{tabular}{|c|c|c|c|c|}
\hline
Model & $n(w_\phi)$ & $\tilde{g}=g/\mphi^{\rm end}$ & $\mu$ \\ \hline
\multirow{5}{*}{E} & 1 (0) & null & null \\ \cline{2-4}
& 2 (1/3) & $4.9\times10^{-28}-5\times10^{-5}$ & $6.0\times10^{-16}-6\times10^{-7}$ \\ \cline{2-4}
 & 3 (0.5) & $ 2.34\times 10^{-34} - 1.22 \times 10^{-6}$ & $8.81 \times 10^{-28} - 8.38 \times 10^{-10}$ \\ \cline{2-4}
 & 6 (5/7) & $8.95 \times 10^{-33} - 7.36 \times 10^{-9}$ & $1.62 \times 10^{-32} - 1.65 \times 10^{-12}$ \\ \cline{2-4}
  & 15 (7/8) & $7.77 \times 10^{-28} - 2.21 \times 10^{-10}$ & $2.1 \times 10^{-31} - 1.65 \times 10^{-14}$ \\ \hline
  \multirow{5}{*}{T} & 1 (0) & null & null \\ \cline{2-4}
  & 2 (1/3) & null & null \\ \cline{2-4}
 & 3 (0.5) & $9.19\times 10^{-34} - 4.36 \times 10^{-20}$ & $1.89 \times 10^{-27} - 7.9 \times 10^{-19}$ \\ \cline{2-4}
 & 6 (5/7) & $7.8 \times 10^{-33} - 2.52 \times 10^{-13}$ & $3.04 \times 10^{-32} - 1.26 \times 10^{-16}$ \\ \cline{2-4}
  & 15 (7/8) & $5.68 \times 10^{-28} - 7.4 \times 10^{-12}$ & $4.34 \times 10^{-31} - 3.48 \times 10^{-16}$ \\ \hline
\end{tabular}
\end{table*}

\begin{figure*}[t!]
         \begin{center}
\includegraphics[width=0017.50cm,height=07.cm]
          {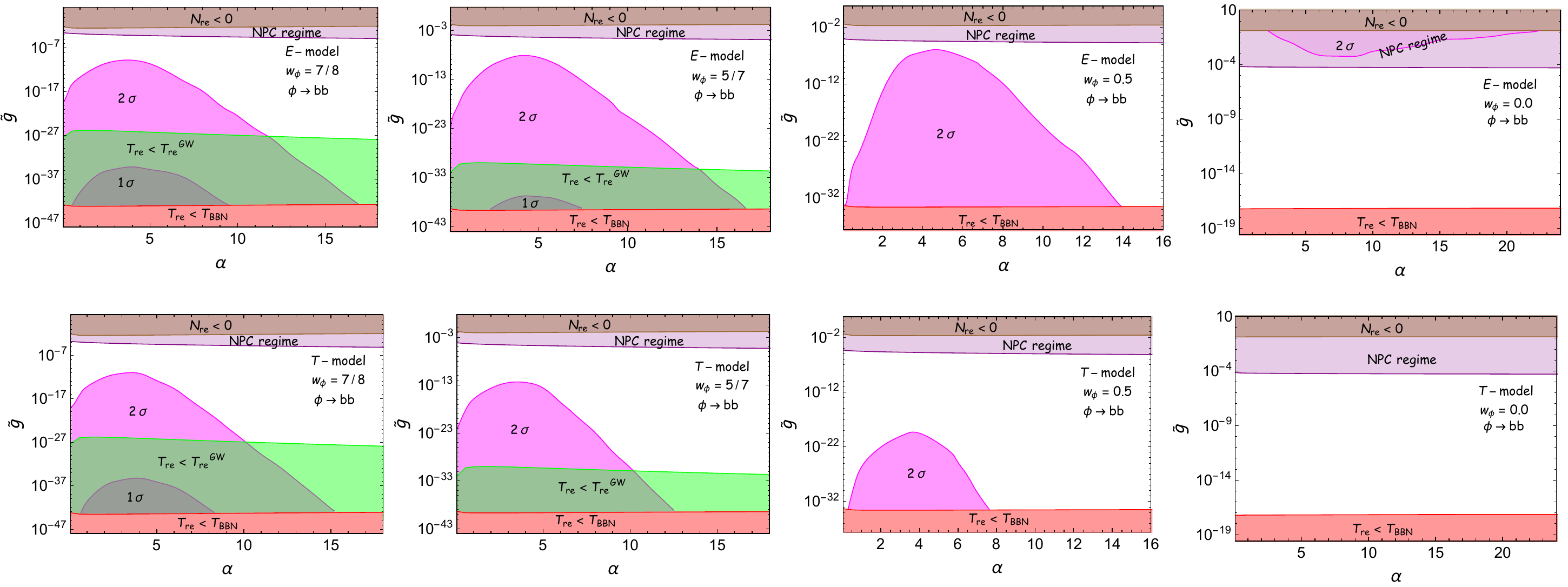}
          \caption{\em The impact of {\rm \textbf{P+ACT+LB+BK18+BBN+PGWs+NPC}} on dimensionless trilinear coupling $\tilde g=g/\mphi^{\rm end}$ with respect to $\alpha$ both for  E-model (upper pannel) and T-model (lower pannel). We consider four representative values of the inflaton EoS $w_{\phi} = (7/8,5/7,0.5,0.0)$. The red and green shaded region indicates $T_{\rm re} < T_{\rm BBN}$ and $T_{\rm re} < T_{\rm re}^{\rm GW}$, respectively, indicating regions excluded by BBN and GW overproduction.  Brown-shaded region implies no reheating ($N_{\rm re}<0$). The purple shaded regions depict NPC regions where the inflaton coupling predictions obtained using the perturbative framework are not reliable, since the perturbative description breaks down in this regime.}
          \label{trilinear}
          \end{center}
      \end{figure*}
\begin{figure*}
         \begin{center}       \includegraphics[width=0017.50cm,height=8.0cm]{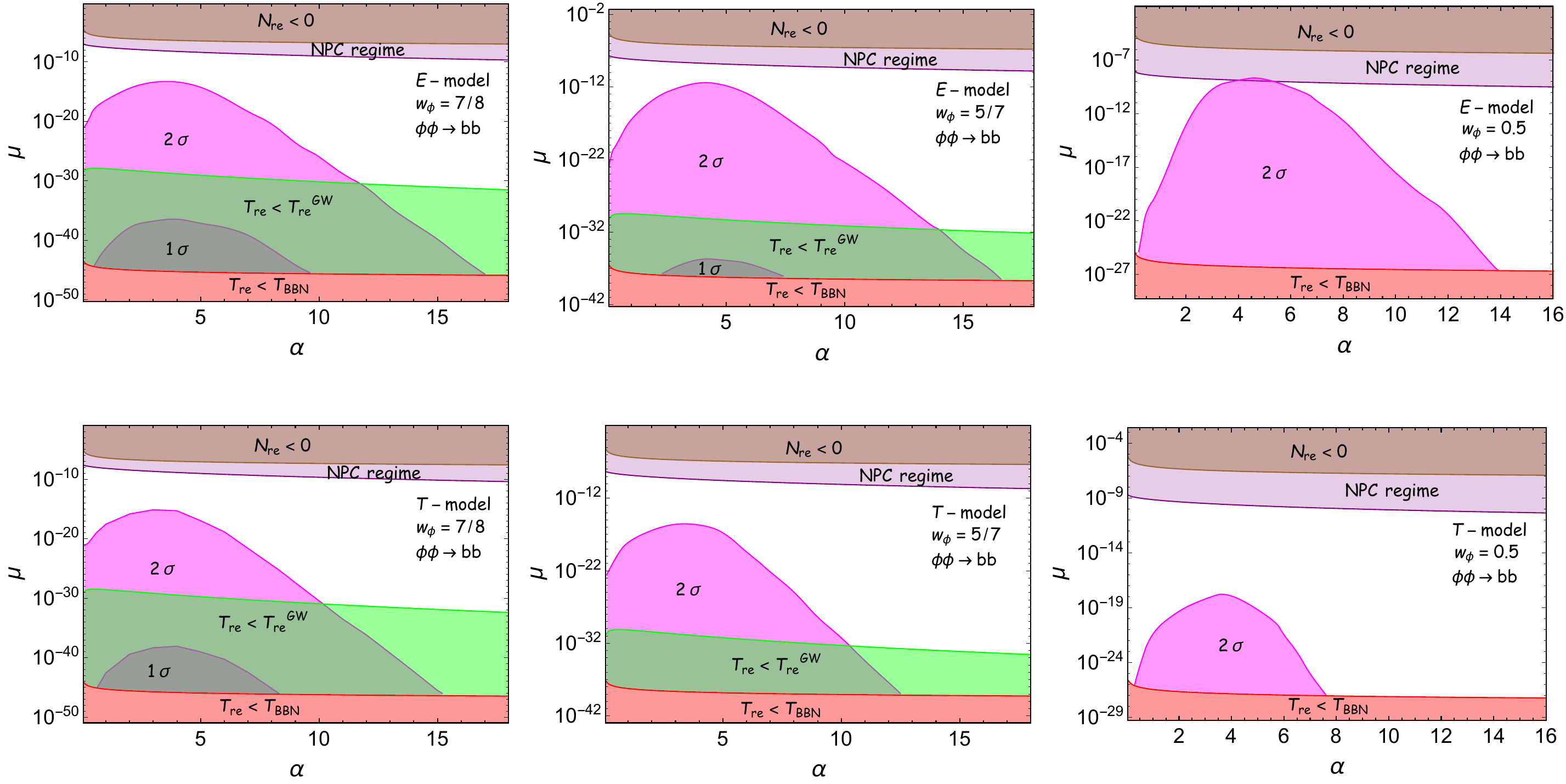}

          \caption{\em The impact of {\rm \textbf{P+ACT+LB+BK18+BBN+PGWs+NPC}} on dimensionless quartic coupling $\mu$ with respect to $\alpha$ both for  E-model (upper pannel) and T-model (lower pannel). The remaining description follows that of Fig.~\ref{trilinear}.} 
          \label{quartic}
          \end{center}
      \end{figure*}
\subsection{Model Constraints}
\subsubsection{E-model}
We have presented our theoretical model predictions of $(n_s, r)$ for various combinations of the potential parameter $\alpha$ and the inflaton EoS parameter $\wphi$ in the $(\ns-r)$ plane, as shown in Fig.~\ref{nsrE} (upper panel). These predictions span reheating temperatures ranging from the maximum possible reheating temperature $\tre^{\rm max} \sim 10^{15}$ GeV (solid black lines) down to the BBN threshold $\tre^{\rm min}=\tbbn \sim 4$ MeV (dashed black lines). In Fig.~\ref{nsrE}, the magenta shaded regions represent the combined observational constraints from P+ACT+LB+BK18. For comparison, the earlier Planck results \cite{Tristram:2021tvh} are shown as blue shaded regions. In this work, we study the constraints on inflationary and post-inflationary parameters using the latest combined dataset P+ACT+LB+BK18. We consider five representative values of the inflaton equation-of-state parameter, $\wphi = (0,\,1/3,\,1/2,\,5/7,\,7/8)$, the corresponding allowed ranges of the potential parameter $\alpha$, together with the inflationary parameters $H_{\rm end}$ and $N_k$, are summarized in Table~\ref{alphatable}. From Fig.~\ref{nsrE}, one can clearly see how the reheating temperature bounds translate into constraints on the parameter $\alpha$. For $\wphi<1/3$, the lower reheating temperature limit $\tre^{\rm min}=\tbbn$ always lies outside the ACT-allowed region in the $(n_s,r)$ plane. As a result, the upper reheating temperature limit $\tre^{\rm max}$ provides the relevant constraint, which in turn sets the maximum allowed value of the potential parameter $\alpha$. In contrast, for $\wphi > 1/3$, the $\tre^{\rm max}$ line always lies outside the ACT contour. In this case, the lower reheating temperature bound-either $\tbbn$ or $T_{\rm re}^{\rm GW}$ (whichever is stronger)-determines the allowed range of $\alpha$. But, for the special case $\omega = 1/3$, the entropy-conservation relation (Eq.~\ref{entropy-conservation}), further reduced to $61.6 \simeq N_k + \ln\!\left(\frac{\rho_{\rm end}^{1/4}}{H_k}\right)$ \cite{Cook:2015vqa}. This relation is independent of the reheating temperatures and predicts a single value of $n_s$ for a given $\wphi$ and $\alpha$ (see Fig.~\ref{w13}).

In Fig.~2, we observe that for a given value of $\alpha$, the full reheating temperature range $(4\,{\rm MeV} \leq T_{\rm re} \leq 10^{15}\,{\rm GeV})$ does not lie entirely within the ACT-allowed region in the $(n_s,r)$ plane. For relatively small values of $\alpha$, the predictions corresponding to the entire reheating temperature range lie outside the ACT contour. As $\alpha$ increases, a portion of these predictions moves inside the ACT-allowed region, so that only part of the reheating temperature range becomes compatible with the observational constraints. However, when $\alpha$ becomes large, the predictions again lie outside the ACT contour. Therefore, only an intermediate range of $\alpha$ allows reheating temperatures that are consistent with ACT observations. To clearly illustrate how the reheating temperature allowed by ACT depends on the potential parameter $\alpha$, we translate the information from the $(n_s,r)$ plane into the $T_{\rm re}$--$\alpha$ plane.

Therefore, in Fig.~\ref{Tree} (upper panel), we present the predicted reheating temperature as a function of the potential parameter $\alpha$ for the $\alpha$-attractor E-model. The different colored regions represent various physical constraints. The red and green shaded areas correspond to $T_{\rm re} < T_{\rm BBN}$ and $T_{\rm re} < T_{\rm re}^{\rm GW}$, respectively, and are thus excluded due to conflicts with BBN and GWs overproduction. The brown-shaded region corresponds to $\tre>10^{15}\,(N_{\rm re}<0)$, implies no reheating. The magenta shaded regions show the predicted reheating temperature obtained from the combined observational constraints at the $1\sigma$ and $2\sigma$ confidence levels. The predicted $1\sigma$ and $2\sigma$ regions are considered viable only if the regions lie outside the excluded regions mentioned above. From the figure, one can clearly see that the allowed reheating temperature strongly depends on the value of the inflaton EoS parameter $w_\phi$. For example, when $w_\phi=0$, the $1\sigma$ predicted values of $T_{\rm re}$ lie entirely above the maximum reheating temperature $T_{\rm re}^{\rm max}\sim10^{15}\,{\rm GeV}$. Consequently, no viable $1\sigma$ region appears in this case (see the upper rightmost panel of Fig.~\ref{Tree}). A similar situation occurs for $w_\phi=0.5$, where the predicted $1\sigma$ reheating temperature lies below the BBN bound $T_{\rm re}^{\rm min}\sim4\,{\rm MeV}$, and therefore the entire $1\sigma$ region is excluded. For larger values of the EoS, such as $w_\phi=5/7$ and $7/8$, the predicted $1\sigma$ reheating temperatures lie above the BBN bound but fall below the stronger constraint $T_{\rm re}^{\rm GW}$. As a result, these regions are ruled out by the PGWs constraint. Again, in the case $w_\phi=1/3$, the predicted scalar spectral index $n_s$ lies outside the $1\sigma$ observational contour (see Fig.~\ref{w13}) for all values of $\alpha$. Consequently, we do not obtain any viable $1\sigma$ region for the reheating temperature in our analysis. 

At the $2\sigma$ level, however, the parameter space remains viable. For a matter-like EoS ($\wphi=0.0$), the combined data from P+ACT+LB+BK18 allows a range of $\tre = (10^{15}-10^{11.6})$ GeV within the $2\sigma$ confidence level, highlighted in magenta. As $w_\phi$ increases, the model predictions for $(n_s,r)$ shift towards the right side of the plane (see Fig.~\ref{nsrE}), resulting in better agreement with the observational constraints. Consequently, the reheating temperature allowed by the observations increases and a wider range of $T_{\rm re}$ becomes compatible with the data. For example, for $w_\phi=0.5$, the allowed reheating temperature spans a broad range, from $T_{\rm re}\sim10^{12.7}\,{\rm GeV}$ down to the BBN bound $T_{\rm re}\sim4\,{\rm MeV}$. For the stiff EoS with $\wphi \gtrsim 0.6$, constraints from the overproduction of PGWs become important;  for $w_\phi=5/7$ and $7/8$, some part of the $2\sigma$ predicted region is excluded by the $T_{\rm re}^{\rm GW}$ bound (green shaded regions). All regions lying outside the $2\sigma$ contour are excluded. The allowed ranges of $\tre$ for different $\wphi$ values, based on the combined constraints from P+ACT+LB+BK18+BBN+PGWs, are summarized in Table~\ref{alphatable}. Finally, the purple shaded regions depict NPC regions, the reheating temperature predicted by ACT observations within this region is not excluded; however, the corresponding inflaton coupling predictions obtained using the perturbative framework are not reliable, since the perturbative description breaks down in this regime. For $\wphi=0$, the reheating temperature $\tre$ allowed by ACT observations lies entirely within the NPC region (see the upper rightmost panel). But, for higher  $\wphi=0.5,5/7$ and $7/8$, all the allowed $2\sigma$ prediction are lie entirely within perturbative regions.

From the particle physics perspective, it is crucial to derive constraints on the microphysical inflaton-SM coupling parameters. Since the inflaton-SM coupling directly determines the reheating temperature, the constraint on $T_{\rm re}$ directly provides bounds on the coupling parameter, as shown in Fig.~\ref{trilinear}, \ref{quartic}. All the color regions have the same significance as we already mentioned. For lower values of EoS, $\wphi < 1/3$, the parameter space allowed by ACT observations predominantly lies in the non-perturbative regime. For the case $\wphi=0$ with bosonic decay process, the reheating temperature $\tre$ allowed by ACT observations lies entirely within the NPC region. Therefore, the corresponding inflaton coupling predictions obtained using the perturbative reheating framework are not reliable in this case (see the upper rightmost panel of Fig.~\ref{trilinear}). In contrast, for higher values EoS, $\wphi>1/3$, the parameter space mostly lies within the perturbative regime. Therefore, within the context of the $\alpha$-attractor E-model, ACT observations favor the non-perturbative reheating for lower EoS $\wphi<1/3$, whereas for higher EoS ($w_\phi>1/3$), ACT observations favor the perturbative reheating. Since we have used a perturbative reheating framework to derive the coupling constraints, the allowed ranges of the perturbative coupling parameters consistent with the ACT observations are summarized in Table~\ref{couplingtable}.

\subsubsection{T-model}
We have performed a similar analysis for the $\alpha$-attractor T-model and presented the corresponding results in Figs.~\ref{nsrE},~\ref{Tree}, \ref{trilinear}, and \ref{quartic} (see all lower panel). Unlike the E-model, the T-model with $\wphi=0.0\,, 1/3$ is not favorable with the recent combined observation P+ACT+LB+BK18 (see Figs.\ref{nsrE} and \ref{w13}). However, for larger values of $\wphi$ such as $0.5, 5/7,$ and $7/8$, certain regions of the parameter space are allowed by observations, similar to the E-model. Despite this similarity, the allowed parameter ranges differ significantly between the two models. For instance, for $\wphi = 0.5$, the maximum allowed value of $\alpha$ for the T-model is 7.8, whereas for the E-model it extends up to 13.9. This indicates that the T-model permits comparatively smaller values of the $\alpha$ parameter. Likewise, the reheating temperature predictions are also different. For $\wphi = 0.5$, the maximum allowed reheating temperature for the T-model is of the order $\mathcal{O}(10^{5})$ GeV, while for the E-model it reaches as high as $\mathcal{O}(10^{12})$ GeV. Since the reheating temperature predictions differ between the models, the corresponding predictions for the microphysical inflaton-SM coupling are also different (see Table-\ref{couplingtable}).

\section{Gravitational dark matter: Constraining its mass}{\label{sc8}}
Gravitational freeze-in production of DM is universal in nature and hence will always be present in any inflationary scenario. We have considered both bosonic dark matter (BDM) and fermionic dark matter (FDM). The perturbative gravitational production rates are,
\begin{eqnarray}{\label{rx}}
    && R_{\rm DM}=\left\{\begin{array}{ll}
         & \frac{\rho^2_{\rm \phi}}{8\pi M^4_{\rm p}}\Sigma^{0}_{\rm b}\,,~~~~~~~~~~~~~~~\mbox{BDM} \\
         & \frac{\rho^2_{\rm \phi}}{2\pi M^4_{\rm p}}\frac{m^2_{\rm f}}{\gamma^2 m^2_{\rm\phi}}\Sigma^{1/2}_{\rm f}~~~~~~~~\mbox{FDM}
    \end{array}\right.
\end{eqnarray}
where $m_{\rm b,f}$ is the DM mass and 
\begin{equation}
\begin{aligned}
&\Sigma^{0}_{\rm b}= \sum^{\infty}_{\nu =1}\lvert\mathcal P^{(2n)}_\nu \rvert^2\left(1+\frac{2m^2_{\rm b}}{\nu^2\gamma^2m^2_{\rm\phi}}\right)^{2}\left(1-\frac{4m^2_{\rm b}}{\nu^2\gamma^2m^2_{\rm\phi}}\right)^{1/2}\,,\\
  &\Sigma^{1/2}_{\rm f}= \sum^{\infty}_{\nu=1}\frac{1}{\nu^2}\lvert\mathcal P^{(2n)}_\nu \rvert^2\left(1-\frac{4m^2_{\rm f}}{\nu^2\gamma^2m^2_{\rm\phi}}\right)^{3/2}.\nonumber
\end{aligned}
\end{equation}
Here, the Fourier coefficients \( \mathcal{P}_\nu^{(2n)} \) are defined through the harmonic expansion of the inflaton potential $V(\phi)$,
\begin{eqnarray} 
V(\phi)=V(\phi_0)\,\mathcal{P}^{2\,n}(t) = V(\phi_0)\sum_{\nu=-\infty}^{\infty} \mathcal{P}^{(2n)}_\nu e^{i\,\nu\,\omega\, t}. 
\end{eqnarray}
The evolution of DM number density ($n_{\rm DM}$) is governed by the Boltzmann equation
\be
\dot{n}_{\rm {DM}}+3Hn_{ \rm DM}=R_{\rm DM}\,, \label{numx}
\ee
\begin{figure*}[t]
         \begin{center}
\includegraphics[width=0015.50cm]
          {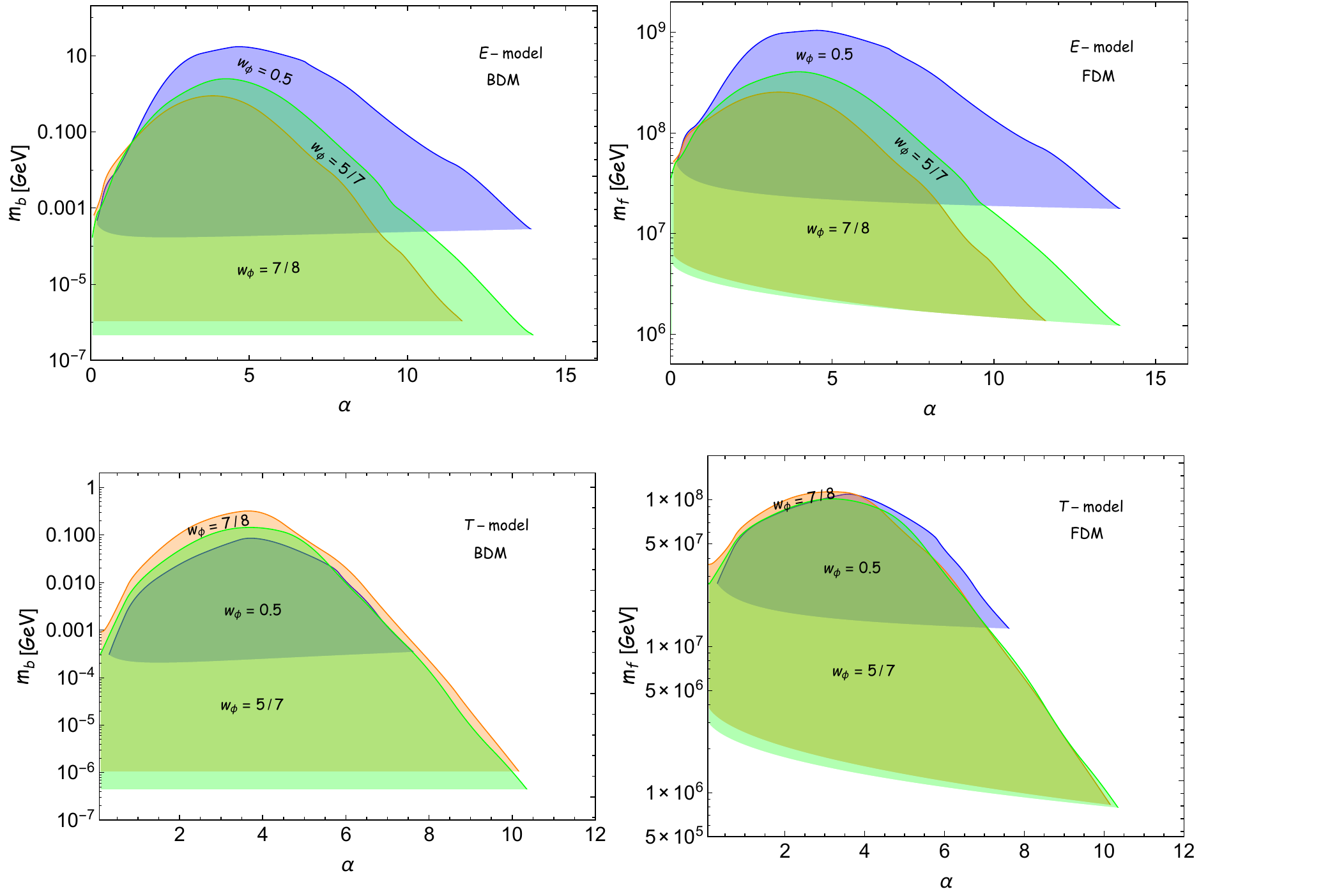}
          \caption{\em Constraints on the gravitational DM mass within the allowed limit of $\alpha$ and $\tre$ for different $\wphi=0.5\,,5/7\,,7/8$.The left (right) panel shows results for bosonic (fermionic) dark matter. The upper and lower panels correspond to the $\alpha$-attractor E-model and T-model, respectively.}
          \label{dmpara}
          \end{center}
      \end{figure*}
      \begin{table*}[ht]\label{DM table}
\centering
\caption{Bounds on the DM mass from  {\rm P+ACT+LB+BK18+BBN+PGWS} for the $\alpha$-attractor model.}
\label{DMtable}
\begin{tabular}{|c|c|c|c|c|}
\hline
Model & $n (w_\phi)$ & $m_{\rm b} (\rm {GeV})$ & $m_{\rm f} (\rm {GeV})$  \\ \hline
\multirow{3}{*}{E} 

 & 3 (0.5) & $ 2.0\times 10^{-4} - 1.679 \times 10^{1}$ & $1.72 \times 10^{7} - 1.03 \times 10^{9}$ \\ \cline{2-4}
 & 6 (5/7) & $4.42 \times 10^{-7} - 2.52 \times 10^{0}$ & $7.92 \times 10^{5} - 9.84 \times 10^{7}$ \\ \cline{2-4}
  & 15 (7/8) & $1.02 \times 10^{-6} - 8.9 \times 10^{-1}$ & $7.94 \times 10^{5} - 1.09 \times 10^{8}$ \\ \hline
  \multirow{3}{*}{T} 
  
 & 3 (0.5) & $2.0\times 10^{-4} - 8.0 \times 10^{-2}$ & $1.3 \times 10^{7} - 1.1 \times 10^{8}$ \\ \cline{2-4}
 & 6 (5/7) & $4.54 \times 10^{-7} - 1.5 \times 10^{-1}$ & $7.82 \times 10^{5} - 1.05 \times 10^{8}$ \\ \cline{2-4}
  & 15 (7/8) & $1.01 \times 10^{-6} - 3.0 \times 10^{-1}$ & $7.98 \times 10^{5} - 1.27 \times 10^{8}$ \\ \hline
\end{tabular}
\end{table*}
Using the Eqs.~(\ref{rx}) in Eq.(\ref{numx}), we have obtained the following solutions for the number
density at the end of reheating $\are$
\begin{eqnarray}
    &&n_{\rm DM}(\are)\simeq\left\{\begin{array}{cc}
         \frac{3H^3_{end}\Sigma^{0}_{\rm b}}{4\pi(1+3w_\phi)}\left(\frac{\are}{a_{\rm end}}\right)^{-3}\,,~~~\mbox{BDM}\\
         \frac{3H^3_{end}m^2_{{\rm f}}\Sigma^{1/2}_{\rm f}}{\pi(1-w_\phi)(\gamma\, m^{\rm end}_\phi)^2}\left(\frac{\are}{a_{\rm end}}\right)^{-3}.~~~\mbox{FDM}
    \end{array}\right.
\end{eqnarray}
The present-day DM relic abundance $\Omega_{\rm DM}h^2= \Omega_{\rm R}h^2\frac{m_{\rm x}n_{\rm x}(a_{\rm re})}{\epsilon T_{\rm now}T^3_{\rm re}}$ can be written as
\begin{eqnarray}
    &&\Omega_{\rm DM}h^2
    \simeq\left\{\begin{array}{ll}
   \Omega_{\rm R}h^2 \frac{3\Sigma^{0}_{\rm b}}{4\pi\epsilon(1+3w_\phi)}\frac{H^3_{end}m_{{\rm b}}}{T_{\rm now}T^3_{\rm re}}(\frac{a_{\rm re}}{a_{\rm end}})^3~~\mbox{BDM}\,,\\
\Omega_{\rm R}h^2 \frac{3\Sigma^{1/2}_{\rm f}}{\pi\epsilon(1-w_\phi)}\frac{m^2_{\rm f}}{(\gamma m^{\rm end}_{\rm\phi})^2}\frac{m_{\rm f}H^3_{end}}{T_{\rm now}T^3_{\rm re}}(\frac{a_{re}}{a_{\rm end}})^3~\mbox{FDM}.
    \end{array}\right.
\end{eqnarray}
In terms of reheating temperature $T_{\rm re}$, the DM abundance can be written as,
\begin{eqnarray}
\frac{\Omega_{\rm DM}h^2}{\Omega_{\rm R}h^2}\simeq\left\{\begin{array}{ll}
      \frac{3H^3_{end}m_{{\rm b}}\Sigma^{0}_{\rm b}}{4\pi\epsilon(1+3w_\phi)T_{\rm now}}\left(\frac{\epsilon}{3M^2_{\rm p}H^2_{\rm end}}\right)^{\frac{1}{1+w_{\rm\phi}}}T_{\rm re}^{\frac{1-3w_{\rm\phi}}{1+w_{\rm\phi}}}\,,\\
      \frac{3H^3_{end}m^3_{{\rm f}}\Sigma^{1/2}_{\rm f}}{\pi\epsilon(1-w_\phi)(\gamma m^{\rm end}_\phi)^2T_{\rm now}}\left(\frac{\epsilon}{3M^2_{\rm p}H^2_{\rm end}}\right)^{\frac{1}{1+w_{\rm\phi}}}T_{\rm re}^{\frac{1-3w_{\rm\phi}}{1+w_{\rm\phi}}}.
\end{array}\right.
\end{eqnarray}
Since DM is produced from the inflaton, its abundance depends only on the reheating temperature and is insensitive to the detailed evolution of the thermal history. As a result, both bosonic decay and bosonic annihilation channels yield identical expressions for the DM abundance. Interestingly, in the case of gravitational DM production, the abundance depends only on the DM mass for given values of $w_\phi$ and $\tre$, and is independent of any arbitrary inflaton-DM coupling. 

In the previous section, we discussed in detail the impact of the latest observational constraints from P+ACT+LB+BK18 on inflation and reheating. Since the DM abundance depends on both the inflationary parameter ($H_{\rm end}$) and the reheating temperature ($\tre$), we explore the allowed DM mass range within the constrained inflation and reheating parameter space. The results are shown in Fig.~\ref{dmpara}, where we identify the DM mass range that yields the correct relic abundance today, $\Omega_{\rm DM} h^2 = 0.118$ \cite{ACT:2025fju,ACT:2025tim}, consistent with the recent ACT DR6 observations.

In Fig.~\ref{dmpara}, we present the allowed DM mass ranges for both bosonic (left panel) and fermionic (right panel) DM candidates, considering three representative values of the inflaton EoS parameter: $\wphi = 0.5$ (blue), $5/7$ (green), and $7/8$ (orange). Since matter-like perturbative reheating ($\wphi = 0$) is disfavored by ACT observations in the case of bosonic decay, and is altogether not viable for bosonic annihilation, no DM predictions can be made for this scenario. Furthermore, for the T-model, $\wphi = 1/3$ is also disfavored by the combined observational data, leading to a null DM prediction in this case as well. However, for the E-model, $\wphi = 1/3$ remains observationally consistent. Interestingly, for $\wphi = 1/3$, the DM abundance becomes independent of the reheating temperature $\tre$. In this case, the correct relic abundance is satisfied for a fixed bosonic DM mass $m_{\rm b} \sim \mathcal{O}(10^2)$ GeV, and for fermionic DM, $m_{\rm f} \sim \mathcal{O}(10^9)$ GeV. Table-\ref{DMtable} summarizes the allowed DM mass ranges for both E-model and T-model scenarios. A comparative analysis reveals that the T-model generally allows for a lower DM mass compared to the E-model.
\section{Conclusions}\label{sc9}
In this work, we have performed a comprehensive phenomenological analysis of the latest combined cosmological data set (P+ACT+LB+BK18) on inflation and the subsequent reheating phase. We examined the $\alpha$-attractor inflationary framework in conjunction with two distinct reheating scenarios: (i) perturbative decay of the inflaton into SM bosons via $\phi \rightarrow b\,b$, and (ii) bosonic annihilation $\phi\,\phi \rightarrow b\,b$. In addition to observational constraints, we incorporated the bound of $\Delta N_{\rm eff}$ arising from the overproduction of PGWs, which constrain the reheating dynamics by constraining the reheating temperature. Furthermore, we included non-perturbative constraints (NPC) on the inflaton-SM couplings, ensuring the validity of the perturbative Boltzmann treatment.

At the first step, considering the most conservative bound of reheating temperature ($T_{\rm BBN}\,\text{to}\,10^{15}\,\rm GeV$), we project our result in the ($\ns-r$) plane to compare with the ACT observation and fix the upper limit on the potential parameter $\alpha$. Once we fixed $\alpha$ parameter, that in turn fixes inflationary parameters such as the range of $N_k$, $H_{\rm end}$, and also post-inflationary parameter like reheating temperature $\tre$. The limiting values of inflationary parameters and reheating temperature for different inflation models are shown in Table-\ref{alphatable}. \\
However, from a particle physics perspective, it is crucial to derive constraints on the microphysical inflaton’s coupling, which plays a key role in the energy transfer process from the inflaton to radiation. Therefore, in the second step, we analyze the constraints on the inflaton’s couplings $\tilde{g}\,,\mu$, taking into account the bounds on inflationary parameters and the reheating temperature. The allowed limit corresponding to the perturbative couplings for different $\wphi$ are shown in Table-\ref{couplingtable}.\\
Finally, we also explored bounds on DM masses produced via universal gravitational production during the reheating era. We considered both bosonic and fermionic DM candidates, whose relic abundances are sensitive to the inflationary and reheating parameters. The resulting mass constraints differ between the E-model and T-model scenarios, as summarized in Table~\ref{DMtable}.

\section{acknowledgments}
RM and AC would like to thank the Ministry of Human Resource Development, Government of India (GoI), for financial assistance. SM extends sincere gratitude to the Department of Science and Technology (DST) for their generous financial support through the INSPIRE scheme (Fellowship No IF190758). 
\appendix
\section{Constraining the inflaton-SM coupling without BE effect}\label{CIP}
\begin{figure*}[t!]
         \begin{center}
\includegraphics[width=0017.50cm,height=07.0cm]
          {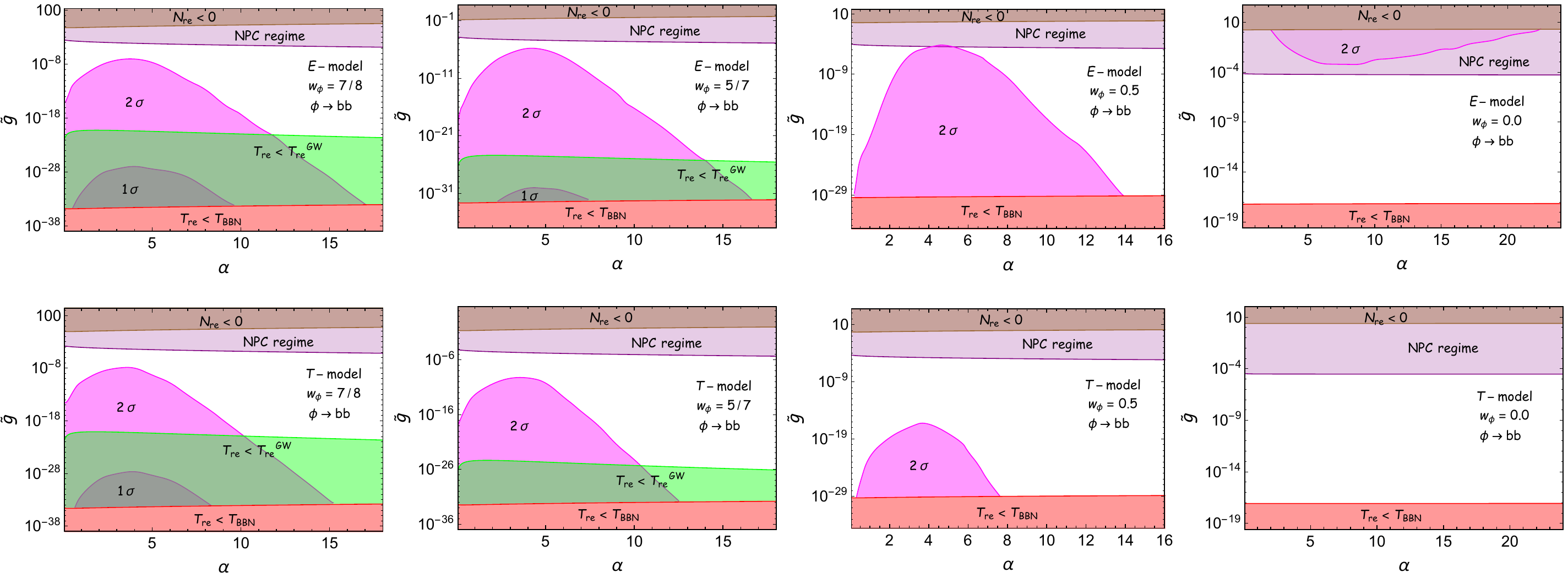}
          \caption{ Description of the plot is the same as Fig.~\ref{trilinear}, except that the BE effect is neglected here}
          \label{wbetrilinear}
          \end{center}
      \end{figure*}

\begin{figure*}
         \begin{center}       \includegraphics[width=0017.50cm,height=7.0cm]{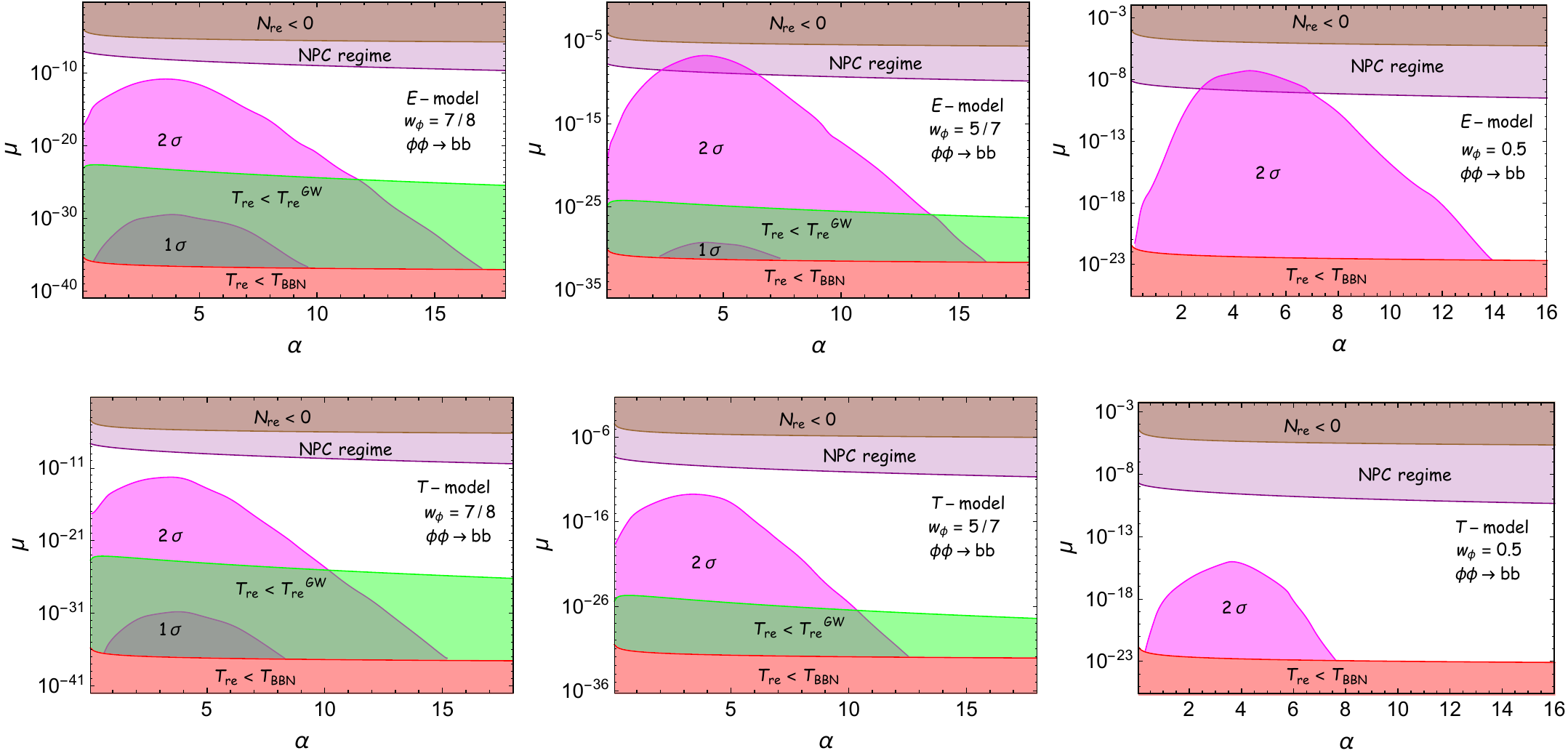}

          \caption{Description of the plot is the same as Fig.~\ref{quartic}, except that the BE effect is neglected here} 
          \label{wbequartic}
          \end{center}
      \end{figure*}

In this appendix, we have shown how the trilinear and quartic couplings are constrained by the recent ACT DR6 observations when quantum effects such as Bose enhancement are neglected. Since the predictions for $\ns$ and $r$ do not depend on the details of the thermal history (decay process of inflaton), but only on the magnitude of the reheating temperature and the inflaton EoS, all model predictions remain the same except for the inflaton coupling. The BE effect enhances the particle production rate from the inflaton condensate; its inclusion leads to a smaller predicted value for the coupling parameter compared to the case without the BE effect (see Figs.~\ref{wbetrilinear},\,\ref{wbequartic}). The numerical values of the coupling parameters are summarized in Table-V.
\begin{table}[h!]
\centering
\scriptsize{\caption{Bounds on the inflaton-SM coupling parameter $\tilde{g}\,,\mu$ for the $\alpha$-attractor model without BE effect}}
\label{wbecouplingtable}
\begin{tabular}{|c|c|c|c|c|}
\hline
Model & $n(w_\phi)$ & $\tilde{g}=g/\mphi^{\rm end}$ & $\mu$ \\ \hline
\multirow{5}{*}{E} & 1 (0) & null & null \\ \cline{2-4}
& 2 (1/3) & $2.0\times10^{-23}-5\times10^{-5}$ & $1.50\times10^{-13}-6\times10^{-7}$ \\ \cline{2-4}
 & 3 (0.5) & $  6.72 \times10^{-30} -  3.34\times 10^{-5}$ & $1.83 \times 10^{-23} - 1.18 \times 10^{-9}$ \\ \cline{2-4}
 & 6 (5/7) & $  2.12 \times10^{-26} - 1.75\times 10^{-6}$ & $1.34\times  10^{-26} - 3.43 \times 10^{-9}$ \\ \cline{2-4}
  & 15 (7/8) & $  5.32 \times10^{-22} -   8.84 \times10^{-8}$ & $2.68\times  10^{-25} - 1.65 \times 10^{-11}$ \\ \hline
  \multirow{5}{*}{T} & 1 (0) & null & null \\ \cline{2-4}
  & 2 (1/3) & null & null \\ \cline{2-4}
 & 3 (0.5) & $7.36\times 10^{-29} - 4.73 \times 10^{-17}$ & $1.25 \times 10^{-23} - 1.29 \times 10^{-15}$ \\ \cline{2-4}
 & 6 (5/7) & $8.46 \times 10^{-26} - 7.12 \times 10^{-10}$ & $5.12 \times 10^{-27} - 2.26 \times 10^{-13}$ \\ \cline{2-4}
  & 15 (7/8) & $1.26 \times 10^{-21} - 1.37 \times 10^{-8}$ & $8.92 \times 10^{-26} - 6.5 3\times 10^{-13}$ \\ \hline
\end{tabular}
\end{table}
\hspace{0.5cm}
\newpage
\bibliographystyle{apsrev4-1}
\bibliography{Rajeshreference}

\begin{thebibliography}{99}
\bibitem{Guth:1982ec}
A.~H.~Guth and S.~Y.~Pi,
Phys. Rev. Lett. \textbf{49}, 1110-1113 (1982)
\bibitem{Albrecht:1982mp}
A.~Albrecht, P.~J.~Steinhardt, M.~S.~Turner and F.~Wilczek,
Phys. Rev. Lett. \textbf{48}, 1437 (1982)
\bibitem{Starobinsky:1982ee}
A.~A.~Starobinsky,
Phys. Lett. B \textbf{117}, 175-178 (1982)
\bibitem{Sofue:2000jx}
Y.~Sofue and V.~Rubin,
Ann. Rev. Astron. Astrophys. \textbf{39}, 137-174 (2001)
doi:10.1146/annurev.astro.39.1.137
[arXiv:astro-ph/0010594 [astro-ph]].
\bibitem{CMB-S4:2016ple}
K.~N.~Abazajian \textit{et al.} [CMB-S4],
[arXiv:1610.02743 [astro-ph.CO]].
\bibitem{Planck:2018jri}
Y.~Akrami \textit{et al.} [Planck],
Astron. Astrophys. \textbf{641}, A10 (2020)
[arXiv:1807.06211 [astro-ph.CO]].
\bibitem{BICEP2:2018kqh}
P.~A.~R.~Ade \textit{et al.} [BICEP2 and Keck Array],
Phys. Rev. Lett. \textbf{121}, 221301 (2018)
[arXiv:1810.05216 [astro-ph.CO]].
\bibitem{BICEP:2021xfz}
P.~A.~R.~Ade \textit{et al.} [BICEP and Keck],
Phys. Rev. Lett. \textbf{127}, no.15, 151301 (2021)
[arXiv:2110.00483 [astro-ph.CO]].
\bibitem{Bezrukov:2007ep}
F.~L.~Bezrukov and M.~Shaposhnikov,
Phys. Lett. B \textbf{659}, 703-706 (2008)
[arXiv:0710.3755 [hep-th]].
\bibitem{Ellis:2013nxa}
J.~Ellis, D.~V.~Nanopoulos and K.~A.~Olive,
JCAP \textbf{10}, 009 (2013)
[arXiv:1307.3537 [hep-th]].
\bibitem{Kallosh:2014laa}
R.~Kallosh, A.~Linde and D.~Roest,
JHEP \textbf{09}, 062 (2014)
[arXiv:1407.4471 [hep-th]].
\bibitem{Kallosh:2013hoa}
R.~Kallosh and A.~Linde,
JCAP \textbf{07}, 002 (2013)
[arXiv:1306.5220 [hep-th]].
\bibitem{Kallosh:2013yoa}
R.~Kallosh, A.~Linde and D.~Roest,
JHEP \textbf{11}, 198 (2013)
[arXiv:1311.0472 [hep-th]].
\bibitem{Maity:2019ltu}
D.~Maity and P.~Saha,
Class. Quant. Grav. \textbf{36}, 045010 (2019)
[arXiv:1902.01895 [gr-qc]].
\bibitem{Garcia:2020wiy}
M.~A.~G.~Garcia, K.~Kaneta, Y.~Mambrini and K.~A.~Olive,
JCAP \textbf{04}, 012 (2021)
[arXiv:2012.10756 [hep-ph]].
\bibitem{Garcia:2020eof}
M.~A.~G.~Garcia, K.~Kaneta, Y.~Mambrini and K.~A.~Olive,
Phys. Rev. D \textbf{101}, no.12, 123507 (2020)
[arXiv:2004.08404 [hep-ph]].
\bibitem{Haque:2023yra}
M.~R.~Haque, D.~Maity and R.~Mondal,
[arXiv:2301.01641 [hep-ph]].
\bibitem{Giudice:2000ex}
G.~F.~Giudice, E.~W.~Kolb and A.~Riotto,
Phys. Rev. D \textbf{64}, 023508 (2001)
[arXiv:hep-ph/0005123 [hep-ph]].
\bibitem{Maity:2018dgy}
D.~Maity and P.~Saha,
Phys. Rev. D \textbf{98}, no.10, 103525 (2018)
[arXiv:1801.03059 [hep-ph]].
\bibitem{Haque:2019prw}
M.~R.~Haque and D.~Maity,
Phys. Rev. D \textbf{99}, no.10, 103534 (2019)
[arXiv:1902.09491 [hep-th]].

\bibitem{Haque:2020zco}
M.~R.~Haque, D.~Maity and P.~Saha,
Phys. Rev. D \textbf{102}, no.8, 083534 (2020)
[arXiv:2009.02794 [hep-th]].
\bibitem{Mambrini:2021zpp}
Y.~Mambrini and K.~A.~Olive,
Phys. Rev. D \textbf{103}, no.11, 115009 (2021)
[arXiv:2102.06214 [hep-ph]].

\bibitem{Haque:2021mab}
M.~R.~Haque and D.~Maity,
Phys. Rev. D \textbf{106}, no.2, 023506 (2022)
[arXiv:2112.14668 [hep-ph]].

\bibitem{Clery:2021bwz}
S.~Clery, Y.~Mambrini, K.~A.~Olive and S.~Verner,
Phys. Rev. D \textbf{105}, no.7, 075005 (2022)
[arXiv:2112.15214 [hep-ph]].

\bibitem{Haque:2022kez}
M.~R.~Haque and D.~Maity,
Phys. Rev. D \textbf{107}, no.4, 043531 (2023)
[arXiv:2201.02348 [hep-ph]].

\bibitem{Ellis:2021kad}
J.~Ellis, M.~A.~G.~Garcia, D.~V.~Nanopoulos, K.~A.~Olive and S.~Verner,
Phys. Rev. D \textbf{105}, no.4, 043504 (2022)
[arXiv:2112.04466 [hep-ph]].
\bibitem{Drewes:2022nhu}
M.~Drewes and L.~Ming,
[arXiv:2208.07609 [hep-ph]].
\bibitem{Drewes:2023bbs}
M.~Drewes, L.~Ming and I.~Oldengott,
[arXiv:2303.13503 [hep-ph]].
\bibitem{Grishchuk:1974ny}
L.~P.~Grishchuk,
Zh. Eksp. Teor. Fiz. \textbf{67}, 825-838 (1974)
\bibitem{Starobinsky:1979ty}
A.~A.~Starobinsky,
JETP Lett. \textbf{30}, 682-685 (1979)
\bibitem{Clarke:2020bil}
T.~J.~Clarke, E.~J.~Copeland and A.~Moss,
JCAP \textbf{10}, 002 (2020)
[arXiv:2004.11396 [astro-ph.CO]].
\bibitem{rgi}
M.~Fairbairn, L.~Lopez Honorez and M.~H.~G.~Tytgat,
Phys. Rev. D \textbf{67}, 101302 (2003)
[arXiv:hep-ph/0302160 [hep-ph]].

\bibitem{Drewes:2017fmn}
M.~Drewes, J.~U.~Kang and U.~R.~Mun,
JHEP \textbf{11}, 072 (2017)
[arXiv:1708.01197 [astro-ph.CO]].
\bibitem{Kawasaki:2000en}
M.~Kawasaki, K.~Kohri and N.~Sugiyama,
Phys. Rev. D \textbf{62}, 023506 (2000)
[arXiv:astro-ph/0002127 [astro-ph]].
\bibitem{Hannestad:2004px}
S.~Hannestad,
Phys. Rev. D \textbf{70}, 043506 (2004)
[arXiv:astro-ph/0403291 [astro-ph]].
\bibitem{Dai:2014jja}
L.~Dai, M.~Kamionkowski and J.~Wang,
Phys. Rev. Lett. \textbf{113}, 041302 (2014)
[arXiv:1404.6704 [astro-ph.CO]].
\bibitem{Cook:2015vqa}
J.~L.~Cook, E.~Dimastrogiovanni, D.~A.~Easson and L.~M.~Krauss,
JCAP \textbf{04}, 047 (2015)
[arXiv:1502.04673 [astro-ph.CO]].
\bibitem{LIGOScientific:2016jlg}
B.~P.~Abbott \textit{et al.} [LIGO Scientific and Virgo],
Phys. Rev. Lett. \textbf{118}, no.12, 121101 (2017)
[erratum: Phys. Rev. Lett. \textbf{119}, no.2, 029901 (2017)]
[arXiv:1612.02029 [gr-qc]].
\bibitem{Punturo:2010zz}
M.~Punturo, M.~Abernathy, F.~Acernese, B.~Allen, N.~Andersson, K.~Arun, F.~Barone, B.~Barr, M.~Barsuglia and M.~Beker, \textit{et al.}
Class. Quant. Grav. \textbf{27}, 194002 (2010)
\bibitem{Crowder:2005nr}
J.~Crowder and N.~J.~Cornish,
Phys. Rev. D \textbf{72}, 083005 (2005)
[arXiv:gr-qc/0506015 [gr-qc]].
\bibitem{Seto:2001qf}
N.~Seto, S.~Kawamura and T.~Nakamura,
Phys. Rev. Lett. \textbf{87}, 221103 (2001)
[arXiv:astro-ph/0108011 [astro-ph]].
\bibitem{LISA:2017pwj}
P.~Amaro-Seoane \textit{et al.} [LISA],
[arXiv:1702.00786 [astro-ph.IM]].
\bibitem{Janssen:2014dka}
G.~Janssen, G.~Hobbs, M.~McLaughlin, C.~Bassa, A.~T.~Deller, M.~Kramer, K.~Lee, C.~Mingarelli, P.~Rosado and S.~Sanidas, \textit{et al.}
PoS \textbf{AASKA14}, 037 (2015)
[arXiv:1501.00127 [astro-ph.IM]].
\bibitem{Mishra:2021wkm}
S.~S.~Mishra, V.~Sahni and A.~A.~Starobinsky,
JCAP \textbf{05}, 075 (2021)
[arXiv:2101.00271 [gr-qc]].
\bibitem{Haque:2021dha}
M.~R.~Haque, D.~Maity, T.~Paul and L.~Sriramkumar,
Phys. Rev. D \textbf{104}, no.6, 063513 (2021)
[arXiv:2105.09242 [astro-ph.CO]].
\bibitem{Vagnozzi:2020gtf}
S.~Vagnozzi,
Mon. Not. Roy. Astron. Soc. \textbf{502}, no.1, L11-L15 (2021)
[arXiv:2009.13432 [astro-ph.CO]].
\bibitem{Benetti:2021uea}
M.~Benetti, L.~L.~Graef and S.~Vagnozzi,
Phys. Rev. D \textbf{105}, no.4, 043520 (2022)
[arXiv:2111.04758 [astro-ph.CO]].
\bibitem{Planck:2018vyg}
N.~Aghanim \textit{et al.} [Planck],
Astron. Astrophys. \textbf{641}, A6 (2020)
[erratum: Astron. Astrophys. \textbf{652}, C4 (2021)]
[arXiv:1807.06209 [astro-ph.CO]].
\bibitem{Coleman:1973jx}
S.~R.~Coleman and E.~J.~Weinberg,
Phys. Rev. D \textbf{7}, 1888-1910 (1973)
\bibitem{Drees:2021wgd}
M.~Drees and Y.~Xu,
JCAP \textbf{09}, 012 (2021)
[arXiv:2104.03977 [hep-ph]].
\bibitem{Kofman:1997yn}
L.~Kofman, A.~D.~Linde and A.~A.~Starobinsky,
Phys. Rev. D \textbf{56}, 3258-3295 (1997)
[arXiv:hep-ph/9704452 [hep-ph]].
\bibitem{Lozanov:2019jxc}
K.~D.~Lozanov,
[arXiv:1907.04402 [astro-ph.CO]].
\bibitem{Dufaux:2006ee}
J.~F.~Dufaux, G.~N.~Felder, L.~Kofman, M.~Peloso and D.~Podolsky,
JCAP \textbf{07}, 006 (2006)
[arXiv:hep-ph/0602144 [hep-ph]].
\bibitem{Greene:2000ew}
P.~B.~Greene and L.~Kofman,
Phys. Rev. D \textbf{62}, 123516 (2000)
[arXiv:hep-ph/0003018 [hep-ph]].
\bibitem{Greene:1998nh}
P.~B.~Greene and L.~Kofman,
Phys. Lett. B \textbf{448}, 6-12 (1999)
[arXiv:hep-ph/9807339 [hep-ph]].
\bibitem{Greene:1997fu}
P.~B.~Greene, L.~Kofman, A.~D.~Linde and A.~A.~Starobinsky,
Phys. Rev. D \textbf{56}, 6175-6192 (1997)
[arXiv:hep-ph/9705347 [hep-ph]].

\bibitem{Tristram:2021tvh}
M.~Tristram, A.~J.~Banday, K.~M.~G\'orski, R.~Keskitalo, C.~R.~Lawrence, K.~J.~Andersen, R.~B.~Barreiro, J.~Borrill, L.~P.~L.~Colombo and H.~K.~Eriksen, \textit{et al.}
Phys. Rev. D \textbf{105}, no.8, 083524 (2022)
[arXiv:2112.07961 [astro-ph.CO]].
\bibitem{CMB-S4:2020lpa}
K.~Abazajian \textit{et al.} [CMB-S4],
Astrophys. J. \textbf{926}, no.1, 54 (2022)
[arXiv:2008.12619 [astro-ph.CO]].
\bibitem{LiteBIRD:2022cnt}
E.~Allys \textit{et al.} [LiteBIRD],
[arXiv:2202.02773 [astro-ph.IM]].
 \end{thebibliography}
\end{document}